\documentclass[11pt,a4paper]{article}
\pdfoutput=1
\usepackage{jcappub}
\usepackage{bm}
\usepackage{empheq}
\usepackage{graphicx}
\usepackage{subcaption}
\usepackage{wrapfig, framed}

\newcommand{\fnl}{f_{\mathrm{NL}}}
\newcommand{\tnl}{\tau_{\mathrm{NL}}}
\newcommand{\gnl}{g_{\mathrm{NL}}}
\newcommand{\be}{\begin{equation}}
\newcommand{\ee}{\end{equation}}
\newcommand{\bea}{\begin{eqnarray}}
\newcommand{\eea}{\end{eqnarray}}
\newcommand{\eref}{\eqref}

\newcommand{\vect}[1]{\bm{\mathrm{{#1}}}}

\graphicspath{{./}}

\usepackage[T1]{fontenc} 

\usepackage{braket}

\numberwithin{equation}{section}

\newcommand\ben{\begin{enumerate}}
\newcommand\een{\end{enumerate}}
\newcommand\bal{\begin{align*}}
\newcommand\eal{\end{align*}}
\newcommand\bi{\begin{itemize}}
\newcommand\ei{\end{itemize}}

\def\I{\item}

\def\id{\protect{{1 \kern-.28em {\rm l}}}}

\newcommand{\gae}{\lower 2pt \hbox{$\, \buildrel {\scriptstyle >}\over {\scriptstyle
\sim}\,$}}
\newcommand{\lae}{\lower 2pt \hbox{$\, \buildrel {\scriptstyle <}\over {\scriptstyle
\sim}\,$}}

\begin{document}

\title{The Separate Universe Approach to Soft Limits}

\author{Zachary Kenton}
\emailAdd{z.a.kenton@qmul.ac.uk}
\author{and David J. Mulryne}
\emailAdd{d.mulryne@qmul.ac.uk}

\affiliation{School of Physics and Astronomy, Queen Mary University of London,\\
Mile End Road, London, E1 4NS,  UK.}

\date{\today}

\abstract{
We develop a formalism for calculating soft limits of $n$-point inflationary 
correlation functions using separate universe techniques. Our 
method naturally allows for multiple fields and leads to an elegant diagrammatic approach. 
As an application we focus on the trispectrum produced by inflation with multiple light fields, giving explicit 
formulae for all possible single- and double-soft limits. We also investigate consistency relations and 
present an infinite tower of inequalities between soft correlation functions which generalise the 
Suyama-Yamaguchi inequality.
 }

\keywords{Inflation, Non-Gaussianity, Trispectrum, Squeezed Limits, Soft Limits}

\maketitle
\newpage
\tableofcontents
\section{Introduction}

Observational cosmology constrains the correlations of primordial perturbations that we believe were produced during inflation. 
Soft limits of cosmological correlation functions occur when there is a large hierarchy between scales involved in the correlation (here soft means a longer wavelength perturbation) and are interesting both observationally 
and theoretically. 
From the observational point of view, future experiments 
will be able to probe a much 
larger range of scales than is currently available \cite{PhysRevD.77.123514,2012PhRvL.109b1302P}.
We therefore need to be able to calculate correlations between perturbations on 
very different scales in order to compare theories against these observations. 
On the theoretical side, soft limits represent an important simplification to the calculation of correlation functions, leading to elegant analytic expressions and to 
consistency relations which apply for broad classes of models.  

Soft limits come in two types: \textit{squeezed} - where an external wavevector becomes soft, or \textit{collapsed} where an internal wavevector (i.e. a sum of external wavevectors) becomes soft. 

In Maldacena's seminal work \cite{Maldacena:2002vr} he found that the squeezed limit of the bispectrum in single-field slow-roll inflation was determined by the tilt of the power spectrum, providing a consistency relation between these observables. The result was 
found to hold more generally for all single-field models with a Bunch-Davies initial state 
and where the classical solution is an attractor \cite{Creminelli:2004yq,Cheung:2007sv}. 
More general single-field soft limits have subsequently been studied, providing further 
consistency relations amongst correlation functions \cite{2011JCAP...11..038C,Khoury:2008wj,2012JCAP...05..037R,Ashoorioon:2013eia,PhysRevD.74.121301,Mooij:2015yka,Li:2008gg,Leblond:2010yq,Seery:2008ax,Hinterbichler:2012nm,
Creminelli:2012ed,McFadden:2014nta,Hinterbichler:2013dpa,
Weinberg:2003sw,Senatore:2012wy,Goldberger:2013rsa,Weinberg:2005vy,Senatore:2009cf,
Flauger:2013hra,Pimentel:2013gza,Berezhiani:2014tda,Tanaka:2011aj,Pajer:2013ana,
Creminelli:2011sq,Berezhiani:2014kga,Binosi:2015obq}. Multiple-soft limits (involving more than one soft mode) 
were considered for single-field inflation in \cite{Joyce:2014aqa,Mirbabayi:2014zpa}.

Soft limits have also been shown to be sensitive to additional fields present 
during inflation \cite{Assassi:2012zq,Kehagias:2015jha,Arkani-Hamed:2015bza,Mirbabayi:2015hva,Sugiyama:2011jt,Sugiyama:2012tr,PhysRevLett.107.191301,PhysRevD.77.023505}. 
In our earlier work \cite{1507.08629}, we considered the case of inflation driven by multiple light 
fields. Employing separate universe techniques \cite{Lyth:1984gv,Wands2000}
including the $\delta N$ expansion  \cite{Sasaki1995,Lyth2004}, we gave explicit analytic 
expressions for the squeezed limit of the bispectrum in multi-field inflation for the first time.
We found that our expression for the reduced bispectrum in a squeezed 
configuration can be significantly different to the standard expression for the reduced bispectrum 
in a close to equilateral configuration \cite{Lyth2005}  (contrast Eq.~\eref{fnlsqueeze}  with Eq.~\eref{fnlequil}).
Shortly thereafter, Byrnes et al. \cite{1601.01970, 1511.03129} 
applied a similar approach to study the hemispherical asymmetry 
(see e.g. \cite{Erickcek:2008sm,Erickcek:2008jp,Kobayashi:2015qma,Byrnes:2015asa,Kenton:2015jga}) and its 
relation to the squeezed bispectrum and collapsed limit of the trispectrum.

In this paper our aim is to extend our earlier results and to investigate general soft limits in
multi-field inflation utilising expansions similar to the $\delta N$ expansion. 

In particular, we show how to produce soft limit diagrams 
with associated rules that lead to compact expressions for soft limits of the 
correlation functions of $\zeta$. 
Our approach can be applied very generally to multiple-soft limits of squeezed (external) and collapsed (internal) momenta of arbitrary $n$-point cosmological correlation functions for models of inflation 
with any number of fields. In this sense, this paper can be viewed as an extension to the multiple-soft limit results of \cite{Joyce:2014aqa,Mirbabayi:2014zpa}. 
As applications, we apply our approach to explore all the single- and 
multiple-soft limits of the trispectrum, and to derive an infinite tower of inequalities 
between soft limits of correlation functions, generalizing the Suyama-Yamaguchi inequality 
\cite{PhysRevD.77.023505,Assassi:2012zq,PhysRevLett.107.191301,Rodriguez:2013cj} to higher point correlation functions.

The compact expressions obtained from using the Soft Limit Expansion Eq.~\eref{softexpansion1} provide analytic insight, but 
are not easy to evaluate explicitly. This is because the coefficients they contain can only easily 
be calculated in certain circumstances, such as for inflation with multiple light scalar 
fields.  Moreover, they
contain field-space correlations of soft 
perturbations evaluated at a later time than the horizon exit time of the soft perturbations. 
On the other hand the objects which are more easily calculated are
these correlations evaluated at the horizon exit time\footnote{These are far from trivial to calculate, but the statistics are expected to be close to Gaussian for canonical models with light fields, 
and known expressions exist for the two-point \cite{Stewart:1993bc,Nakamura:1996da}, 
three-point \cite{astro-ph/0506056,Gao:2008dt} and four point correlation 
functions \cite{Seery:2006js,Seery:2008ax} in many circumstances.}. We therefore introduce one further separate 
universe expansion -- the $\Gamma$ expansion \cite{Yokoyama:2007uu,Yokoyama:2007dw,Anderson:2012em,Mulryne:2013uka,Seery:2012vj} -- which allows these later 
correlations to be calculated in terms of the horizon crossing correlations, and then present 
explicit expressions for the soft limits of inflation with multiple light scalar fields. 

This paper is laid out as follows.
In \S\ref{sec:deltaN} we review the standard $\delta N$ expansion.
In \S\ref{sec:soft} we consider soft limits. We introduce a new Soft Limit Expansion in \S\ref{sec:softExpansion}, 
which is a form of background wave method used by other authors. As a simple example we calculate 
formal expressions for the squeezed bispectrum and collapsed trispectrum 
in \S\ref{sec:simpleExamples}, showing how the Suyama-Yamaguchi inequality arises in our approach. 
In \S\ref{subseb:softdiag} we introduce the soft limit diagrams. We then use them to quickly 
calculate all other soft limits of the trispectrum in \S\ref{subsec:SoftEx}, and to find an infinite 
tower of inequalities in \S\ref{sec:ineq} which generalise the Suyama-Yamaguchi 
inequality to higher-point correlation functions. 
In \S\ref{sec:gamma} we give more explicit expressions. We introduce the $\Gamma$ expansion in \S\ref{Gamma} applying it to correlation functions in \S\ref{sec:gammacor} and provide explicit examples for multiple light fields in \S\ref{sec:explicit}. We conclude in \S\ref{conc}. In Appendix~\ref{appBwave} we give more details on the background wave method. In Appendix~\ref{app:single} we show how our multi-field double-soft limit reduces to the consistency relation of 
the single field case. In Appendix~\ref{app:gammadiag} we present a diagrammatic approach for the $\Gamma$ expansion.

\section{The $\delta N$ Formalism}
\label{sec:deltaN}

\subsection{The $\delta N$ Expansion}
\label{subsec:deltaNex}
Consider a multi-field model of inflation with $M$ fields, $\phi_A(t_i,\vect{x}) = \phi_A(t_i) + \delta \phi_A(t_i,\vect{x}) \equiv \phi_A^{(i)} + \delta \phi_A^{(i)}(\vect{x})$, 
where the latin uppercase index $A$ runs from $1$ to $M$, and the superscript $(i)$ denotes evaluation at some initial time $t_i$ during inflation, a shorthand we use throughout.

The $\delta N$ expansion, based on the separate universe approach to cosmological 
perturbation theory \cite{Lyth:1984gv,Wands2000}, states that for a given  
mode $\vect{k}$ which is super-horizon, $k>aH$ (where $k \equiv |\vect{k}|$), 
the primordial curvature perturbation, $\zeta$, can be expanded in Fourier space
as \cite{Sasaki1995,Lyth2005,Lyth2004} ($\star$ below denotes convolution)
\begin{align}
\begin{split}
&\zeta_{\vect{k}}  = \delta N_{\vect{k}} = N^{(i)}_{A } {\delta \phi^{(i)}_{A}}_{\vect{k}} + \frac{1}{2}N^{(i)}_{AB}\left [ \delta \phi^{(i)}_{A } \star
\delta \phi^{(i)}_{B}\right ]_{\vect{k}} + \dots
\\
&\text{where } ~~ N^{(i)}_{A} \equiv \frac{\partial N^{(i)}}{\partial \phi^{(i)}_{A }}, \qquad N^{(i)}_{AB} \equiv \frac{\partial^2 N^{(i)}}{\partial \phi^{(i)}_{B }\partial \phi^{(i)}_{A }}\,, ~~ \textrm{ etc.} 
\end{split}\label{deltaN}
\end{align}
and $N^{(i)}$ is the local number of e-folds from an initial flat hypersurface 
at time $t_i$ to a final uniform density slice at some final time, $t_f$, long after horizon crossing 
when we wish to evaluate the properties of $\zeta$. The field perturbations 
are evaluated in the flat gauge. We do not put a time-superscript on $\zeta$ itself to reduce clutter. 
$N^{(i)}$ and its derivatives can be 
calculated with knowledge only of the background cosmology, a result of the separate universe 
approximation.

A few comments are in order. 
We have written this expansion in terms of field perturbations alone. For multiple light fields 
this captures the leading effects. But we note that it can 
easily be generalised to include field velocities if these 
were important. Moreover 
if other degrees of freedom (vector or tensorial perturbations for example)
were important
 we could 
extend the summation over these degrees of freedom as well. At a formal level 
these extensions are always possible, though the explicit calculation of the coefficients (the 
derivatives of $N$) becomes less clear. 
Going beyond a separate universe approximation we could even include the sensitivity of 
$\zeta$ at the later time to gradient terms at the earlier time -- moving therefore to a gradient expansion \cite{PhysRevD.42.3936,Shibata:1999zs,Deruelle:1994iz}. To keep the notation clean throughout this paper, and because our primary concern is 
models with multiple light fields, we will 
indicate a summation over only field space indices using upper case Roman indices, $A,B,...$. 
We will always bear in mind, however, that this set of variables can be formally extended to all relevant degrees of freedom, and so the results, such as our consistency relations, are rather general.

\subsection{Correlations of the Curvature Perturbation}
The objects of primary interest for observations are the $n-$point correlation functions of 
$\zeta$ evaluated at some late
time relevant to observations $t_f$. 
We introduce arbitrary external momenta, $\vect{k_1},\vect{k_2},...,\vect{k_n}$ which we 
will order, without loss of generality, by their magnitudes $k_1\leq k_2\leq ...\leq k_n$, where $k = |\vect{k}|$.
These scales exit the 
horizon, $k = aH(t)$,  at times $t_1,t_2,...,t_n$ respectively, where $t_1\leq t_2 \leq ...\leq t_n$. 
From now on if we drop the time superscript \emph{on any object other than $\zeta$}, it is to be 
understood that the evaluation time should be the exit time of 
the hardest mode, (shortest wavelength), $t_n$, for example $N\equiv N^{(n)}$ and 
$\delta\phi_A \equiv \delta\phi_A^{(n)}$.  This avoids unnecessary clutter of our expressions.

We introduce notation for $n-$point $\zeta$ correlation functions such that
\begin{align}
\langle   \zeta(\vect{k_1})\cdot \cdot \cdot  \zeta(\vect{k_n})\rangle 
&=  G_n(\vect{k_1},...,\vect{k_n})
 (2 \pi)^3 
  \delta(\vect{k_1} + \cdot \cdot \cdot +\vect{k_n})
\end{align}
so that, for example, $G_2$, $G_3$ and $G_4$ 
represent the power spectrum, bispectrum and trispectrum respectively
\begin{align}
G_2(\vect{k_1}, -\vect{k_1}) &= P_{\zeta}({k_1})
\\
G_3(\vect{k_1}, \vect{k_2}, \vect{k_3}) &= B_{\zeta}({k_1},{k_2},{k_3})
\\
G_4(\vect{k_1}, \vect{k_2}, \vect{k_3}, \vect{k_4}) &= T_{\zeta}(\vect{k_1}, \vect{k_2}, \vect{k_3}, \vect{k_4}).
\end{align}

And in a similar manner for $n-$point field space correlation functions we have
\begin{align}
\langle  \delta \phi _{A_1}(\vect{k_1})\cdot \cdot \cdot \delta \phi _{A_p}(\vect{k_p})\rangle 
&=  F_{A_1 \cdot \cdot \cdot A_p }(\vect{k_1},...,\vect{k_p})
 (2 \pi)^3 
  \delta(\vect{k_1} + \cdot \cdot \cdot +\vect{k_p})\,.\label{Ffieldspace}
\end{align}
For the two, three and four point we also employ the conventional symbols
\begin{align}
F_{A B }(\vect{k_1}, -\vect{k_1}) &= \Sigma_{A B }({k_1})\label{sigmadef}
\\
F_{A B C}(\vect{k_1}, \vect{k_2}, \vect{k_3}) &= \alpha_{A B C}({k_1},{k_2},{k_3})
\\
F_{A BCD}(\vect{k_1}, \vect{k_2}, \vect{k_3}, \vect{k_4}) &= T_{ABCD}(\vect{k_1}, \vect{k_2}, \vect{k_3}, \vect{k_4}).
\end{align}

The $\delta N$ expansion then allows us to write the $\zeta$ correlators in terms of the $\delta \phi_A$ correlators. The result for the power spectrum of $\zeta$ was first given in \cite{Sasaki1995}, the bispectrum in \cite{Lyth2005} and 
the trispectrum in \cite{Seery:2006vu,Byrnes:2006vq,Seery:2006js,Seery:2008ax}. 
Higher point $\zeta$ correlators are related to field-space correlators and can be nicely calculated using the diagrammatic presentation of \cite{Byrnes:2007tm}.

\section{Soft Limits} \label{sec:soft}
\subsection{Soft Limit Expansion} \label{sec:softExpansion}
Soft limits occur when there is a hierarchical separation of scales involved in a correlation. 
For any real-space perturbation, $Y$, (for example, $Y$ can stand for $\zeta$ or for $\delta\phi_A$) we can 
consider two contributions to this perturbation. 
One contains only Fourier modes clustered around some hard mode $1/k_{\rm h}$ 
(subscript $\rm h$ for hard), and the other contains only long modes around some 
arbitrary soft mode, $1/k_{\rm s}$ (subscript $\rm s$ for soft), leading to
\begin{align}
\begin{split}
Y(\vect{x}) &\subset Y^{\rm h}(\vect{x})+Y^{\rm s}(\vect{x})
\\
Y^{\rm h}(\vect{x}) &= \int^{k_{\rm h} + \Delta k_{\rm h} }_{k_{\rm h} - \Delta k_{\rm h}} \frac{{\rm d}^3\vect{k}}{(2\pi)^3} e^{i \mathbf{k}.\vect{x}} Y_{\vect{k}}\\
Y^{\rm s}(\vect{x}) &= \int^{ k_{\rm s}+ \Delta k_{\rm s}}_{ k_{\rm s}- \Delta k_{\rm s}} \frac{{\rm d}^3\vect{k}}{(2\pi)^3} e^{i \vect{k}.\vect{x}} Y_{\vect{k}}
\end{split}
\end{align}
where the ranges $\Delta k_{\rm h}$ and $\Delta k_{\rm s}$ are arbitrary, but small.
This means that the Fourier components of $Y^{\rm h}$ only have support on $[k_{\rm h} - \Delta k_{\rm h},k_{\rm h} + \Delta k_{\rm h}]$ and similarly for $Y^{\rm s}$ whose Fourier components only have support on $[k_{\rm s} - \Delta k_{\rm s},k_{\rm s} + \Delta k_{\rm s}]$.

For soft limits it can be argued that the dominant contribution to correlations between hard and soft modes comes from how the soft modes, which exit the horizon at much earlier times, correlate with the shifts that the soft modes cause in the background cosmology felt by the hard modes. 
This is a form of the background wave assumption, which is discussed at length in Appexdix~\ref{appBwave}.
It can be used for any set of scales, but becomes accurate only when the hierarchy is large. In 
this work we implement this assumption by Taylor expanding the value the hard contribution to 
$\zeta$ takes in the background of the soft contribution to the scalar fields, which 
we denote $\zeta^{\rm h}(\vect{x}) \big |_{\rm s}$, about the value it would have 
taken in the absence of soft scalar field modes, denoted $\zeta^{\rm h}$.
The expansion then, in Fourier space for some hard wavevector $\vect{k}$, is
\begin{align}
{\zeta^{\rm h}_{\vect{k} }}\Big|_{{\rm s}}= \zeta^{\rm h} _{\vect{k}} +  \left [ \zeta^{\rm h} _{,A} \star  \delta \phi_A^{\rm s}\right ]_{\vect{k}}  + \frac{1}{2}
\left [ \frac{ \partial^2 \zeta^{\rm h}   } {\partial \phi_{A}\partial \phi_{B}} \star  \delta \phi_A^{\rm s}\star  \delta \phi_B^{\rm s}\right ]_{\vect{k}}+ \dots \,\label{softexpansion1}
\end{align}
(see Appexdix~\ref{appBwave} for a fuller discussion).
We call Eq.~\eref{softexpansion1} the Soft Limit Expansion. 
It can be seen as a form of separate universe expansion, in which the soft modes alter 
the background cosmology in which hard modes exit and subsequently evolve. The hard modes effectively 
feel a different background in different spatial locations. The derivatives in the Taylor expansion 
are taken with respect to \emph{the background fields}, $\phi_A(t)$, a consequence 
of the separate universe approximation.
The time of evaluation of $\delta \phi_A^{\rm s}$ and $\partial/\partial\phi_A$ 
is arbitrary, as long as they are the same as each other. We will always take it to be the 
last exit time of the modes under consideration within a correlation.

In writing Eq.~\eref{softexpansion1}, we have again 
assumed that the perturbed cosmology can be fully defined at some 
given time just in terms of field fluctuations on flat hypersurfaces, which is the case during slow roll, for example.
As for the $\delta N$ expression, however, there is nothing to stop us, at least at a formal level, from extending 
this to include any other degrees of freedom that may be important.
The expansion Eq.~\eref{softexpansion1} could be generalised to include 
spatial gradients to get subleading soft behaviour - this would be the analogy to 
the gradient expansion \cite{PhysRevD.42.3936,Shibata:1999zs,Deruelle:1994iz} which extends 
the usual $\delta N$. We leave this for future work.

In the following sections we will consider soft limits of correlations. 
We will insert the expansion Eq.~\eref{softexpansion1} 
for all modes considered hard, and assume that the dominant contributions will come from Wick contractions 
amongst soft modes themselves, and Wick contractions amongst hard modes themselves, 
but not between soft and hard modes. This is the mathematical version of the assumption 
that the main contribution to the correlations come from how the soft modes correlate with the shifts they cause to the background cosmology which the hard modes experience. This leads to the factorization of the soft limits of correlations into hard sub-processes. 

\subsection{Simple Examples}
\label{sec:simpleExamples}
Before presenting general rules which allow us to generate expressions for arbitrary soft 
limits let us consider two simple examples: the squeezed limit of the bispectrum and the single-soft collapsed limit of the trispectrum.
\subsubsection{The Squeezed Limit of the Bispectum}
As a first simple example of the use of the 
expansions presented above we revisit the squeezed limit of the bispectrum, considered in our earlier work \cite{1507.08629}.
We take 
$\langle \zeta_{\vect{k_1}} \zeta_{\vect{k_2}} \zeta_{\vect{k_3}} \rangle $, with $k_1 \ll k_2 \sim k_3$, 
and employ Eq.~\eref{softexpansion1} to expand ${\zeta^{\rm h}_{\vect{k_2} }}\big|_{{\rm s}}$ and ${\zeta^{\rm h}_{\vect{k_3} }}\big|_{{\rm s}}$ in 
terms of long field perturbations, and insert these into the correlator
\begin{align}
&\lim_{k_1 \ \rm {soft}} \langle \zeta_{\vect{k_1}} \zeta_{\vect{k_2}} \zeta_{\vect{k_3}} \rangle \approx \langle \zeta_{\vect{k_1}}^{\rm s} {\zeta^{\rm h}_{\vect{k_2} }}\big|_{{\rm s}}{\zeta^{\rm h}_{\vect{k_3} }}\big|_{{\rm s}} \rangle
\\
&\approx \langle \zeta_{\vect{k_1}}^{\rm s} 
\left (\zeta^{\rm h} _{\vect{k_2}} +  \left [ \frac{ \partial \zeta^{\rm h}   } {\partial \phi_{A}} \star  \delta \phi_A^{\rm s}\right ]_{\vect{k_2}} + \dots\right )
\left (\zeta^{\rm h} _{\vect{k_3}} +  \left [ \frac{ \partial \zeta^{\rm h}   } {\partial \phi_{B}} \star  \delta \phi_B^{\rm s}\right ]_{\vect{k_3}} + \dots\right )
 \rangle
 \\
&\approx  (2 \pi)^3 
  \delta(\vect{k_1} + \vect{k_2} +\vect{k_3})\langle \zeta_{\vect{k_1}}^{\rm s}\delta {\phi_B^{\rm s}}_{-\vect{k_1}}\rangle'\frac{ \partial    } {\partial \phi_{B}}\left (\frac{1}{2}P_{\zeta}(k_2) + \frac{1}{2}P_{\zeta}(k_3)\right )
\end{align}
where the primed correlator denotes the correlator stripped of the delta function and the factor of $(2 \pi)^3$.
In the first line we make the soft limit assumption that the dominant contribution in the soft limit comes from the correlation between the soft modes and the change that the soft modes cause in the hard modes. This allows us to replace $\zeta_{\vect{k_1}}$ with $\zeta^{\rm s}_{\vect{k_1}}$, and to replace $\zeta_{\vect{k_2}}$ with ${\zeta^{\rm h}_{\vect{k_2} }}\big|_{{\rm s}}$ within the correlator (and similarly for $\zeta_{\vect{k_3}}$).
 In the final line we use Wick's theorem and the soft limit assumption that only soft modes correlate with soft modes, and only hard modes correlate with hard modes. 
 
 Now, in the soft limit to leading order $k_2 \approx k_3$, we have that the final line simplifies to
\begin{align}
\lim_{k_1 \ \rm {soft}} \langle \zeta_{\vect{k_1}} \zeta_{\vect{k_2}} \zeta_{\vect{k_3}} \rangle&\approx  (2 \pi)^3 
  \delta(\vect{k_1} + \vect{k_2} +\vect{k_3}) N_A \Sigma_{AB}(k_1) P_{\zeta ,B} (k_3)
  \label{singlesoft3pt}
\end{align}
where we also used the first order $\delta N$ expansion Eq.~\eref{deltaN} for the soft $\zeta$, and the notation $Y_{,B} \equiv\frac{ \partial Y     } {\partial \phi_{B}} $, for any function $Y$, together with Eq.~\eref{sigmadef} for the definition of $\Sigma_{AB}(k_1)$.

The expression Eq.~\eref{singlesoft3pt} is quite formal, but very compact. In \S\ref{sec:gamma} we will see how to turn it into a more explicit expression which can be evaluated to gain model specific predictions.

\subsubsection{The Collapsed Limit of the Trispectum}
\begin{figure}[htb!]
\begin{framed}
\centering
\begin{subfigure}[t]{0.3\textwidth}
\centering
\includegraphics[scale=0.25]{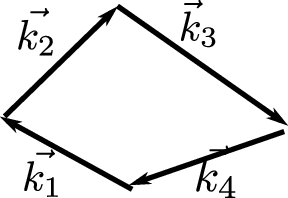} 
\caption{Equilateral }
\end{subfigure}
\begin{subfigure}[t]{0.3\textwidth}
\centering
\includegraphics[scale=0.25]{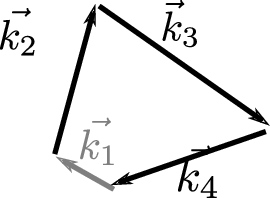} 
\caption{Single-Soft Squeezed}
\end{subfigure}
\begin{subfigure}[t]{0.3\textwidth}
\centering
\includegraphics[scale=0.25]{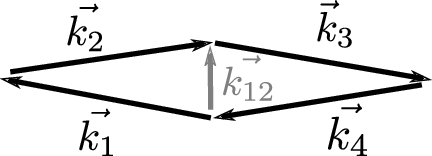} 
\caption{Single-Soft Collapsed}
\end{subfigure}
\begin{subfigure}[t]{0.3\textwidth}
\centering
\includegraphics[scale=0.25]{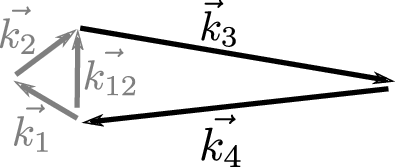} 
\caption{Double-Soft Kite}
\end{subfigure}
\begin{subfigure}[t]{0.3\textwidth}
\centering
\includegraphics[scale=0.25]{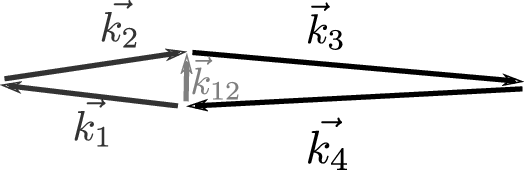} 
\caption{Double-Soft Squished}
\end{subfigure}
\begin{subfigure}[t]{0.3\textwidth}
\centering
\includegraphics[scale=0.25]{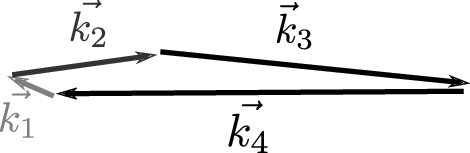} 
\caption{Double-Soft Wonky}
\end{subfigure}
\caption{Possible soft shapes of the trispectrum. For illustration we have drawn the quadrilaterals here as planar, but in general they can be three-dimensional. The greyscale adds emphasis, with lighter grey being more soft, while darker grey is more hard.}
\label{fig:tri_shapes}
\end{framed}
\end{figure}

\noindent The possible soft limit shapes for the trispectrum are shown in Fig.~\ref{fig:tri_shapes}.
As a second example we consider the single-soft collapsed limit of the trispectrum, $k_{12} \ll k_1 \approx k_2 \sim k_3 \approx k_4$, where $\vect{k_{12}} = \vect{k_1} + \vect{k_2}$, illustrated in Fig.~\ref{fig:tri_shapes}(c).
We consider the four point function, 
$\langle \zeta_{\vect{k_1}} \zeta_{\vect{k_2}} \zeta_{\vect{k_3}} \zeta_{\vect{k_4}} \rangle $ with all the external $\zeta$'s taken to be hard with 
respect to the soft collapsed mode $k_{12}$. We use 
Eq.~\eref{softexpansion1} on each of the four $\zeta$'s and insert these into the correlator
\begin{align}
\lim_{k_{12} \  \rm {soft}} \langle & \zeta_{\vect{k_1}} \zeta_{\vect{k_2}} \zeta_{\vect{k_3}} \zeta_{\vect{k_4}}\rangle 
\approx 
\langle  
{\zeta^{\rm h}_{\vect{k_1} }}\big|_{{\rm s}}
{\zeta^{\rm h}_{\vect{k_2} }}\big|_{{\rm s}}
{\zeta^{\rm h}_{\vect{k_3} }}\big|_{{\rm s}}
{\zeta^{\rm h}_{\vect{k_4} }}\big|_{{\rm s}}
 \rangle
\\
\begin{split}
\approx 
\langle  &
\left (\zeta^{\rm h} _{\vect{k_1}} +  \left [ \zeta^{\rm h} _{,A} \star  \delta \phi_A^{\rm s}\right ]_{\vect{k_1}} + \dots\right )
\left (\zeta^{\rm h} _{\vect{k_2}} +  \left [ \zeta^{\rm h} _{,B} \star  \delta \phi_B^{\rm s}\right ]_{\vect{k_2}} + \dots\right )\times
\\
&\left (\zeta^{\rm h} _{\vect{k_3}} +  \left [ \zeta^{\rm h} _{,C} \star  \delta \phi_C^{\rm s}\right ]_{\vect{k_3}} + \dots\right )
\left (\zeta^{\rm h} _{\vect{k_4}} +  \left [ \zeta^{\rm h} _{,D} \star  \delta \phi_D^{\rm s}\right ]_{\vect{k_4}} + \dots\right )
 \rangle.
\end{split}
\end{align}
There are lots of possible terms that can now appear when we Wick contract.
In what follows we only show the terms that contribute at leading order in the soft limit. These are the terms that contain $\Sigma_{AB}(k_{12})$, which occur either when $\zeta^{\rm h} _{\vect{k_1}}$ gets contracted with the ${\zeta^{\rm h} _{,B}}$ inside the convolution $\left [ \zeta^{\rm h} _{,B} \star  \delta \phi_B^{\rm s}\right ]_{\vect{k_2}}$, or similarly, when $\zeta^{\rm h} _{\vect{k_2}}$ gets contracted with the ${\zeta^{\rm h} _{,A}}$ inside the convolution $\left [ \zeta^{\rm h} _{,A} \star  \delta \phi_A^{\rm s}\right ]_{\vect{k_1}}$. This is because the delta function that accompanies these contractions then forces the integrated momentum in the convolution, (which is the momentum of $\delta \phi_B^{\rm s}$ or $\delta \phi_A^{\rm s}$ respectively) to have magnitude $k_{12}$ -- which then appears in $\Sigma_{AB}(k_{12})$. There are four possible such terms giving the following contribution to the correlator
\begin{align}
\begin{split}
&\lim_{k_{12} \  \rm {soft}} \langle  \zeta_{\vect{k_1}} \zeta_{\vect{k_2}} \zeta_{\vect{k_3}} \zeta_{\vect{k_4}}\rangle 
\\
&\approx 
\int_{\vect{p}}\int_{\vect{q}}
\left ( 
\langle \zeta^{\rm h} _{\vect{k_1}} {\zeta^{\rm h}_{,A}} _{\vect{k_2}-\vect{p}} \rangle 
+ 
\langle {\zeta^{\rm h}_{,A}} _{\vect{k_1}-\vect{p}}\zeta^{\rm h} _{\vect{k_2}}  \rangle
\right )
\langle \delta {\phi_A^{\rm s}}_{\vect{p}}\delta {\phi_B^{\rm s}}_{\vect{q}}\rangle
\left ( 
\langle \zeta^{\rm h} _{\vect{k_3}} {\zeta^{\rm h}_{,B}} _{\vect{k_4}-\vect{q}} \rangle 
+ 
\langle {\zeta^{\rm h}_{,B}} _{\vect{k_3}-\vect{q}}\zeta^{\rm h} _{\vect{k_4}}  \rangle
\right ).
\end{split}
\end{align}
Now using $\langle \zeta^{\rm h} _{\vect{k_1}} {\zeta^{\rm h}_{,A}} _{\vect{k_2}-\vect{p}} \rangle = (2 \pi)^3 
  \delta(\vect{k_1} + \vect{k_2} -\vect{p})\frac{1}{2}P_{\zeta}(k_1)_{,A}$, and similarly for the other three two-point functions, we get
  \begin{align}
  \begin{split}
&\lim_{k_{12} \  \rm {soft}} \langle  \zeta_{\vect{k_1}} \zeta_{\vect{k_2}} \zeta_{\vect{k_3}} \zeta_{\vect{k_4}}\rangle 
\\
&\approx (2 \pi)^3 
  \delta(\vect{k_1} + \vect{k_2} +\vect{k_3} + \vect{k_4})\frac{1}{2}\left [P_{\zeta}(k_1) + P_{\zeta}(k_2)\right ]_{,A}\Sigma_{AB}(k_{12})
\frac{1}{2}\left [P_{\zeta}(k_3) + P_{\zeta}(k_4)\right ]_{,B}.
\end{split}
  \end{align}
Now, since in the soft limit to leading order $k_1 \approx k_2$
and $k_3 \approx k_4$,
we can replace $P_{\zeta}(k_2)$ with $P_{\zeta}(k_1)$ and $P_{\zeta}(k_4)$ with $P_{\zeta}(k_3)$ to get our 
final expression
  \begin{align}
  \begin{split}
&\lim_{k_{12} \  \rm {soft}} \langle  \zeta_{\vect{k_1}} \zeta_{\vect{k_2}} \zeta_{\vect{k_3}} \zeta_{\vect{k_4}}\rangle 
\\
&\approx (2 \pi)^3 
  \delta(\vect{k_1} + \vect{k_2} +\vect{k_3} + \vect{k_4}) P_{\zeta}(k_1)_{,A}\Sigma_{AB}(k_{12})
P_{\zeta}(k_3)_{,B}.
\end{split}\label{internalsoft4pt}
  \end{align}
We note that, though presented in a different manner, this agrees with the calculation 
of Byrnes et al. \cite{1511.03129} for the collapsed limit.
  
\subsubsection{Suyama-Yamaguchi Inequality} \label{sec:Suyama-Yamaguchi}

We can use Eq.~\eref{internalsoft4pt} and Eq.~\eref{singlesoft3pt} to directly 
prove the 
soft limit version of the Suyama-Yamaguchi inequality \cite{PhysRevD.77.023505} relating the single-soft collapsed limit of the trispectrum to the squeezed limit of the bispectrum \cite{2011PhRvL.107s1301S,Assassi:2012zq,Rodriguez:2013cj}. We begin by defining the  dimensionless parameters 
\begin{align}
 \tilde{f}_{\rm NL}(k_{12},k_1,k_2) &\equiv \frac{5}{12}  \frac{\lim\limits_{k_{12} \ \rm {soft}}  B_\zeta(k_{12},k_1,k_2)}{P_\zeta(k_{12}) P_\zeta(k_1)} = \frac{5}{12}\frac{N_A \Sigma_{AB}(k_{12}) P_{\zeta } (k_1)_{,B}}{P_\zeta(k_{12}) P_\zeta(k_1)}\,,\label{fnlsq}
  \\
\tilde{\tau}_{\rm NL}(\vect{k_1}, \vect{k_2},\vect{k_3},\vect{k_4}) &\equiv \frac{1}{4}\frac{\lim\limits_{k_{12} \ \rm {soft}}  T_\zeta(\vect{k_1}, \vect{k_2},\vect{k_3},\vect{k_4})}{P_\zeta(k_{12}) P_\zeta(k_1)P_\zeta(k_3)} =\frac{1}{4}\frac{  {P_\zeta}_{,A}(k_1)  \Sigma_{AB}(k_{12}){P_\zeta}_{,B}(k_3) }{ P_\zeta(k_{12}) P_\zeta(k_1)P_\zeta(k_3)}.\label{tnlcoll}
\end{align}
The equalities that follow the definitions use Eq.~\eref{singlesoft3pt} and Eq.~\eref{internalsoft4pt} respectively. 

The soft version of the Suyama-Yamaguchi inequality follows in the special case where $k_3=k_1$. 
In this case the numerator of Eq.~\eref{tnlcoll} becomes $P_{\zeta}(k_1)_{,A}\Sigma_{AB}(k_{12})
P_{\zeta}(k_1)_{,B}$, which can be viewed as the inner product, with respect to 
the metric $\Sigma_{AB}(k_{12})$, of a vector with components $P_{\zeta}(k_1)_{,A}$. 
The numerator of Eq.~\eref{fnlsq} is $N_A \Sigma_{AB}(k_{12}) P_{\zeta } (k_1)_{,B}$ which is 
the inner product of a different vector, $N_A$, with the original vector $P_{\zeta}(k_1)_{,B}$. 
The Cauchy-Schwarz inequality then gives
\begin{align}
[P_{\zeta}(k_1)_{,A}\Sigma_{AB}(k_{12})P_{\zeta}(k_1)_{,B}]
[N_C \Sigma_{CD}(k_{12})N_D ] \geq [N_E\Sigma_{EF}(k_{12})P_{\zeta}(k_1)_{,F} ]^2. \label{cauchyschwarz}
\end{align}
We can now use the $k_3=k_1$ version of Eq.~\eref{tnlcoll} to replace the the first term in the left hand side of Eq.~\eref{cauchyschwarz} in terms of $\tilde{\tau}_{\rm NL}$, and rewrite the second term using
the $\delta N$ expression $P_{\zeta}(k_{12})=N_C \Sigma_{CD}(k_{12})N_D$ . Finally using Eq.~\eref{fnlsq} to 
replace the RHS in terms of $\tilde f_{\rm NL}$, we arrive at
\begin{align}
\tilde{\tau}_{\rm NL} \geq  \left (\frac{5}{6}\tilde{f}_{\rm NL}\right)^2 \label{sytnlfnl}\,.
\end{align}
This is a rather direct proof of this soft limit relation, which to our knowledge 
hasn't appeared before, but which recovers the results of \cite{2011PhRvL.107s1301S,Assassi:2012zq,Rodriguez:2013cj}. In \S\ref{sec:ineq} we will see how this inequality can be generalised to provide relations between higher point functions.

\subsection{Soft Limit Diagrams}
\label{subseb:softdiag}
The examples given so far were sufficiently simple that we could easily take a direct approach using the expansion Eq.~\eref{softexpansion1} and then Wick contracting, using some algebra to get simple final expressions such as Eq.~\eref{singlesoft3pt} and Eq.~\eref{internalsoft4pt}. For soft limits of higher-point correlation functions, however, this approach becomes cumbersome. It proves useful to generate a set of rules which lead to compact final expressions of the form given above. 
This can be readily achieved since for any soft limit the procedure is simply to insert the Soft Limit Expansion Eq.~\eref{softexpansion1} 
for every hard $\zeta$ perturbation that is present. Wick contractions then occur amongst the soft modes themselves, and between the hard modes themselves, but not between soft and hard modes. Soft and hard modes are correlated only through the derivatives of the Soft Limit Expansion, Eq.~\eref{softexpansion1}.  
If there are $N$ soft momenta of the same size, we need to Taylor expand to $N$-th order consistently in both the Soft Limit Expansion, Eq.~\eref{softexpansion1}, and in
the $\delta N$ expansion of the soft $\zeta$
\begin{align}
\zeta^{\rm s} _{\vect{k}}  = N_{A } {\delta\phi_A^{\rm s}}_{\vect{k}}  + \frac{1}{2}N_{AB}\left  [\delta\phi_A^{\rm s}  \star\delta\phi_B^{\rm s} \right ]_{\vect{k}} + \cdots . \,\label{softzetadeltaN}
\end{align}
We can then organise the result in terms of diagrams, which we call Soft Limit Diagrams. 
These diagrams are analogous to the $\delta N$ graphs \cite{Byrnes:2007tm} which represent the Taylor expansion of standard $\delta N$. 
For simplicity we focus here only on tree level\footnote{Although we note that loops could easily be included in our diagrammatic approach.} contributions and capture only leading order behaviour in the soft limit and gradient expansion. 

We now give rules for how to calculate a correlation in which all of the soft momenta are the same hierarchical size. 
It may be helpful to read these rules in combination with the examples which follow, in order to clarify the proceedure.
\begin{enumerate}
\item Identify all soft squeezed (soft external) momenta and put a box around each one. Identify \textit{all} soft collapsed (internal) momenta built from a group of hard external momenta and put a box around each group. At this stage all external momenta should now be in a box. Draw a black vertex on each box.
\I Connect the black vertices by drawing a connected tree diagram with dashed lines. Each dashed line must connect on one end to a black vertex and on the other end to a white vertex. At a black vertex (possibly multiple) dashed lines can connect to a box. At a white vertex dashed lines connect to other dashed lines.
\I Label each dashed line with a distinct field index $A_1,A_2,...$.
\I Ensure momentum conservation at every vertex, which determines the momentum of each dashed line.
\I The two vertex types are assigned the following factors:
\begin{enumerate}
\I Assign a factor $\left [ G_{Q}(\{\vect{k}\})\right ]_{,A_1\cdot \cdot \cdot A_m}$ to each black vertex which connects a box containing $Q \geq 1$ external momenta, $\{\vect{k}\}$,  to $m$ dashed lines with field indices $A_1\cdot \cdot \cdot A_m$, where $m \geq 1$. Note that for $Q=1$, we have $\left [ G_{1}(\vect{k})\right ]_{,A_1\cdot \cdot \cdot A_m} = N_{A_1\cdot \cdot \cdot A_m}$.
\I Assign a factor $F_{A_1\cdot \cdot \cdot A_s}(\vect{p_1},...,\vect{p_s})$ to each white vertex with $s$ dashed lines with incoming momenta $\vect{p_1},...,\vect{p_s}$ and field index $A_1\cdot \cdot \cdot A_s$, where $s \geq 2$.
\end{enumerate} 
\I Each diagram is associated with the mathematical expression obtained by multiplying together all vertex factors. Repeat the above process from stage 2 onwards to generate all distinct connected tree diagrams. $G_n(\vect{k_1},...,\vect{k_n})$ is then obtained by summing over all these diagrams.
\end{enumerate}

If there are consecutive soft momenta, where there are hierarchies amongst the soft momenta, then follow the rules above for the softest in the hierarchy. Then recursively repeat the same rules for the next level up in the hierarchy, to calculate soft limits of correlators sitting within the hard sub-process box(es).

\subsection{Examples Using Diagrams}
\label{subsec:SoftEx}
We now show some examples of soft limits calculated using the diagrammatic approach. First we revisit the calculations of \S\ref{sec:simpleExamples} using the rules presented above to check they reproduce the same answers. Then we consider other soft limits of the trispectrum, such as single-soft squeezed and various double-soft limits. 

\subsubsection{Simple Examples Revisited}
\begin{figure}[htb!]
\begin{framed}
\center
\includegraphics[scale=0.7]{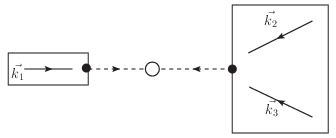} 
\caption{The only tree-level connected soft diagram for the squeezed limit of the bispectrum. }
\label{fig:softbi}
\end{framed}
\end{figure}

\begin{figure}[htb!]
\begin{framed}
\center
\includegraphics[scale=0.7]{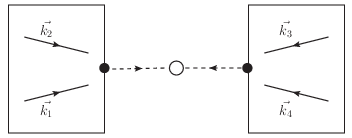} 
\caption{The only tree-level connected soft diagram for the single-soft collapsed limit of the trispectrum. }
\label{fig:tri_s_in}
\end{framed}
\end{figure}

In Fig.~\ref{fig:softbi} we show the diagram one gets for the squeezed limit of the bispectrum. Multiplying the vertex factors together, one can check that this diagram reproduces the soft bispectrum of Eq.~\eref{singlesoft3pt}. 

In Fig.~\ref{fig:tri_s_in} we show the diagram one gets for the single-soft collapsed limit of the trispectrum. Multiplying the vertex factors together, one can check that this diagram reproduces the single-soft collapsed result of Eq.~\eref{internalsoft4pt}.

\subsubsection{Other Examples}

\begin{figure}[htb!]
\begin{framed}
\center
\includegraphics[scale=0.7]{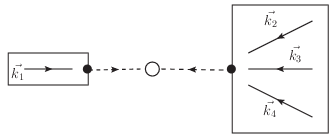} 
\caption{The only tree-level connected soft diagram for the single-soft squeezed limit of the trispectrum.  }
\label{fig:tri_s_ex}
\end{framed}
\end{figure}

We now look at the single-soft squeezed limit of the trispectrum, $k_1 \ll k_2 \approx k_3 \approx k_4$. This has a very similar diagram to the squeezed limit of the bispectrum, and is shown in Fig.~\ref{fig:tri_s_ex}, giving the result 
\begin{align}
\lim _{k_1 \ll k_2 \approx k_3 \approx k_4} T_{\zeta}(\vect{k_1}, \vect{k_2}, \vect{k_3}, \vect{k_4}) = N_A\Sigma_{AB}(k_1)\left [B_{\zeta}(k_2,k_3,k_4)\right ]_{,B}.
\label{singex}
\end{align}

\

Next we consider the double-soft limits of the trispectrum, which in general is given by taking both $k_1 \ll k_3 \approx k_4$ and $k_2 \ll k_3 \approx k_4$. In this double-soft limit we have three choices for how $k_1$, $k_2$ and $k_{12}$ are related, shown in Fig.~\ref{fig:tri_shapes}, which we name
\begin{align}
k_1 \approx k_2 \approx k_{12} &\ll k_3\approx k_4 \qquad \textrm{(kite)} \label{kite}
\\
 k_{12} \ll k_1 \approx k_2   &\ll k_3\approx k_4 \qquad \textrm{(squished)} \label{squished}
 \\
k_1 \ll k_2  &\ll k_3\approx k_4 \qquad \textrm{(wonky)}. \label{wonky}
\end{align}

\noindent \textbf{1. Double-Soft Kite}: $k_1 \approx k_2 \approx k_{12} \ll k_3\approx k_4$

\noindent For the kite shape there are four diagrams which can be drawn, shown in Fig.~\ref{fig:tri_shapes}(d). The sum of the diagrams gives the expression (appearing in relative locations in the expression below)
\begin{align}
\begin{split}
&\lim _{\rm kite} T_{\zeta}(\vect{k_1}, \vect{k_2}, \vect{k_3}, \vect{k_4}) 
\\ = &N_{A}N_{B}P_{\zeta}(k_3)_{,C}\alpha_{ABC}(k_1,k_2,k_{12}) + N_{A}N_{B}P_{\zeta}(k_3)_{,CD}\Sigma_{AC}(k_1)\Sigma_{BD}(k_2)
\\
+& N_{A}N_{BC}P_{\zeta}(k_3)_{,D}\Sigma_{AB}(k_1)\Sigma_{CD}(k_{12})+ N_{A}N_{BC}P_{\zeta}(k_3)_{,D}\Sigma_{AB}(k_2)\Sigma_{CD}(k_{12}).
\end{split}\label{kitetri}
\end{align}

Note that in order to get the bottom line of Eq.~\eref{kitetri} we had to expand $\zeta^{\rm s}_{\vect{k_1}}$ and $\zeta^{\rm s}_{\vect{k_2}}$ to second order in $\delta N$ expansion, using Eq.~\eref{softzetadeltaN}, so as to work to second order consistently throughout the calculation, i.e. we need to work to second order in both the $\delta N$ expansion and the Soft Limit Expansion.
\begin{figure}[htb!]
\begin{framed}
\center
\includegraphics[scale=0.59]{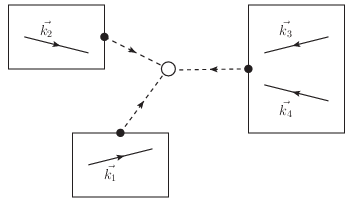} 
\includegraphics[scale=0.59]{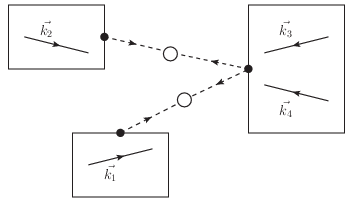} 
\includegraphics[scale=0.59]{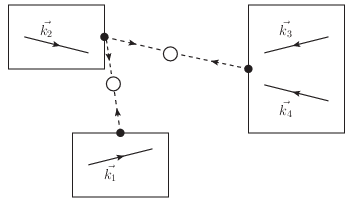}
\includegraphics[scale=0.59]{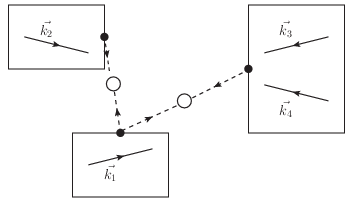} 
 \caption{The four distinct tree-level connected soft diagrams for the double-soft limit of the trispectrum for the kite shape.}
\label{fig:kitediag}
\end{framed}
\end{figure}

\

\noindent \textbf{2. Double-Soft Squished}: $ k_{12} \ll k_1 \approx k_2   \ll k_3\approx k_4$

\noindent For the squished shape we have the same diagram as in Fig.~\ref{fig:tri_s_in}, leading to the same expression as Eq.~\eref{internalsoft4pt}
  \begin{align}
  \begin{split}
&\lim_{k_{12} \  \rm {soft}} \langle  \zeta_{\vect{k_1}} \zeta_{\vect{k_2}} \zeta_{\vect{k_3}} \zeta_{\vect{k_4}}\rangle 
\\
&\approx (2 \pi)^3 
  \delta(\vect{k_1} + \vect{k_2} +\vect{k_3} + \vect{k_4}) P_{\zeta}(k_1)_{,A}\Sigma_{AB}(k_{12})
P_{\zeta}(k_3)_{,B}.
\end{split}\label{squished2}
  \end{align}

\

\noindent \textbf{3. Double-Soft Wonky}: $k_1 \ll k_2  \ll k_3\approx k_4 $

\noindent For the wonky shape we have to use the soft diagram rules recursively. The soft diagram is shown in Fig.~\ref{fig:tri_shapes}(f). The diagram is constructed in two steps. First one draws a  box around $k_1$, which is the softest momentum, and another box around the other three momenta. The second step, inside the box containing the other three momenta, is to draw a sub box around $k_2$, which is the next softest momenta, and another around the remaining two momenta. The resulting expression is
\begin{align}
\lim _{\rm wonky} T_{\zeta}(\vect{k_1}, \vect{k_2}, \vect{k_3}, \vect{k_4}) 
 = &N_{A}\Sigma_{AB}(k_1)\left [N_{C}\Sigma_{CD}(k_2)\left [P_{\zeta}(k_3)\right ]_{,D} \right ]_{,B}.
 \label{wonkytri}
\end{align}
Note that we could expand out the derivatives here, and obtain three terms corresponding to the first three terms of Eq.~\eref{kitetri}, but with $\alpha_{ABC}(k_1,k_2,k_{12})$ replaced by its soft limit counterpart $\lim_{k_1\ll k_2}\alpha_{ABC}(k_1,k_2,k_{12})\approx \Sigma_{AD}(k_1)\Sigma_{BC,D}(k_2)$ \cite{1507.08629}. The fourth term in Eq.~\eref{kitetri}, corresponding to the bottom right diagram of Fig.~\ref{fig:kitediag}, is not present in the wonky limit, as it will be subdominant in the wonky limit compared to the other terms. 

\begin{figure}[htb!]
\begin{framed}
\center
\includegraphics[scale=0.7]{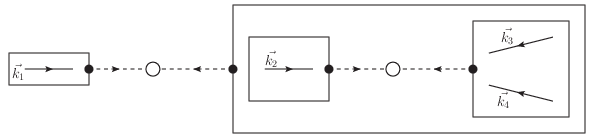} 
\caption{The tree-level connected soft diagram for the wonky shape double-soft limit of the trispectrum.}
\label{fig:wonkydiag}
\end{framed}
\end{figure}

In this section we only considered examples up to the trispectrum for simplicity, but the rules can be applied to higher $n$-point function examples. 
We note that this procedure goes beyond what is available in the current literature, and in particular it allows for multiple  fields. The resulting expressions provide relations between soft limits of $n$-point $\zeta$ correlation functions and derivatives of lower-point $\zeta$ correlation functions, contracted with soft field-space correlation functions.

\subsection{Inequalities Between Soft Correlation Functions}
\label{sec:ineq}

Following our considerations in \S\ref{sec:Suyama-Yamaguchi} of the Suyama-Yamaguchi inequality \cite{PhysRevD.77.023505}, Eq.~\eref{sytnlfnl}, we will now 
show how to generalise this inequality to higher-point functions. In the equilateral limit, and for Gaussian fields, \cite{Suyama:2011qi} found inequalities amongst higher-point functions using the Cauchy-Schwarz inequality. Here we find inequalities for soft limit higher-point functions and without assuming Gaussian fields.

Consider a single-soft limit of an $n$-point function, with momenta ordered such that $k_1 \leq ... \leq k_n$. We take the soft momentum to be $\vect{p} \equiv \sum^r_{i=1} \vect{k_i}$,
where $r=1$ corresponds to a squeezed soft limit, while $2 \leq r \leq n -1$ corresponds to a collapsed soft limit. 
We will take the hard momenta to have wavenumbers all approximately of the size $k_*$.
 We then
apply the soft diagram rules to find
\begin{align}
 \lim_{p \rm \ soft}G_n(\vect{k_1},...,\vect{k_r}, \vect{k_{r+1}},...,\vect{k_n}) \approx \left [G_r(\vect{k_1},...,\vect{k_r})\right ]_{,A}\Sigma_{AB}(p)\left [G_{n-r}(\vect{k_{r+1}},...,\vect{k_n})\right ]_{,B}\,.
 \label{Gnsplitr}
 \end{align} 
Next we identify the following two vectors 
\begin{eqnarray}
X_A &\equiv& \left [G_r(\vect{k_1},...,\vect{k_r})\right ]_{,A}\\
Y_B &\equiv& \left [G_{n-r}(\vect{k_{r+1}},...,\vect{k_n})\right ]_{,B}\,.
\end{eqnarray}
We note that since $\Sigma_{AB}(p)$ is a real symmetric matrix it 
provides an inner product on the vector space in which $X$ and $Y$ live,  
which means we can utilise the Cauchy-Schwarz inequality for $X$ and $Y$
\begin{align}
\left [X_A \Sigma_{AB}(p) Y_B\right ]^2 \leq \left [X_C \Sigma_{CD}(p) X_D\right ]\left [Y_E \Sigma_{EF}(p) Y_F\right ].
\label{CauchySchwarz}
\end{align}
The LHS of Eq.~\eref{CauchySchwarz} gives Eq.~\eref{Gnsplitr}, and since $G_r(\vect{k_1},...,\vect{k_r}) = G_{r}(-\vect{k_{1}},...,-\vect{k_r})$, the RHS of Eq.~\eref{CauchySchwarz} can be written as the product of two other soft limits
\begin{align}
 \lim_{p \rm \ soft}G_{2r}(\vect{k_1},...,\vect{k_r}, -\vect{k_1},...,-\vect{k_r}) &\approx \left [G_r(\vect{k_1},...,\vect{k_r})\right ]_{,A}\Sigma_{AB}(p)\left [G_{r}(-\vect{k_{1}},...,-\vect{k_r})\right ]_{,B} 
 \label{G2rsplitr}
 \\
  \lim_{p \rm \ soft}G_{2(n-r)}(\vect{k_{r+1}},...,\vect{k_n}, -\vect{k_{r+1}},...,-\vect{k_n}) &\approx \left [G_n(\vect{k_{r+1}},...,\vect{k_n})\right ]_{,A}\Sigma_{AB}(p) \nonumber \\
 &\times \left [G_{r}(-\vect{k_{r+1}},...,-\vect{k_n})\right ]_{,B}
 \label{G2(n-r)splitn-r}
 \end{align} 
 which yields
\begin{align}
\left [  \lim_{p \rm \ soft}G_n(\vect{k_1},...,\vect{k_r}, \vect{k_{r+1}},...,\vect{k_n})\right ]^2 
&\leq \left [ \lim_{p \rm \ soft}G_{2r}(\vect{k_1},...,\vect{k_r}, -\vect{k_1},...,-\vect{k_r})\right ] \nonumber \\
& \times \left [   \lim_{p \rm \ soft}G_{2(n-r)}(\vect{k_{r+1}},...,\vect{k_n}, -\vect{k_{r+1}},...,-\vect{k_n})\right ]\,.
 \end{align}
This can be written in terms of soft limit dimensionless parameters
\begin{align}
f_{n}(\vect{k_1},...,\vect{k_r}, \vect{k_{r+1}},...,\vect{k_n}) \equiv
 \lim_{p \rm \ soft}
\frac{ G_n(\vect{k_1},...,\vect{k_r}, \vect{k_{r+1}},...,\vect{k_n})
}{P_{\zeta}(p)[P_{\zeta}(k_*)]^{n-2}}\label{dimensionless}
\end{align}
as
 \begin{align}
 \left [ f_n(\vect{k_1},...,\vect{k_r}, \vect{k_{r+1}},...,\vect{k_n})\right ]^2 &\leq \left [f_{2r}(\vect{k_1},...,\vect{k_r}, -\vect{k_1},...,-\vect{k_r})\right ] \nonumber \\
 &\times \left [  f_{2(n-r)}(\vect{k_{r+1}},...,\vect{k_n}, -\vect{k_{r+1}},...,-\vect{k_n})\right ]
 \end{align}
or, suppressing the momentum dependence for brevity,
\begin{align}
f_n^2 \leq f_{2r} f_{2(n-r)}.
\end{align}
We note that $f_2=1$ by definition, while $f_3 = \frac{12}{5}\tilde{f}_{\rm NL}$ and $f_4 = 4\tilde{\tau}_{\rm NL}$. If we fix $n=3$ and $r=1$, 
one recovers the soft limit Suyama-Yamaguchi inequality, $\left (6 \tilde f_{\textrm{NL}}/5\right )^2\leq \tilde \tau_{\textrm{NL}}$ \cite{PhysRevD.77.023505,2011PhRvL.107s1301S,Assassi:2012zq,Rodriguez:2013cj}. 

Novel, yet similar, relations exist between the single-soft squeezed (external) limit of an $n$-point correlator and the single-soft collapsed (internal) limit of a $2n-2$ correlation function, which follows by fixing $r=1$ and $n>3$. 

For $1 <r < n-1$,
a qualitatively different kind of relation emerges between the single-soft collapsed limit of an $n$-point correlator and the product of a single-soft collapsed limit of a $2r$-point correlator and a single-soft collapsed limit of a $2(n-r)$-point correlator, i.e. all the soft limits in this case are collapsed ones. The first non-trival example of this type of inequality occurs for $n=5$, $r=2$, which relates the collapsed $5$-point function to the collapsed $6$-point function and collapsed $4$ point function. 

Note that in the case of single-source inflation, all of the inequalities are saturated, since the Cauchy-Schwarz inequality becomes an equality in a vector space of only one dimension.

\section{Explicit Expressions}
\label{sec:gamma}

\subsection{The $\Gamma$ expansion}
\label{Gamma}

The soft limit expressions we generated in \S\ref{sec:soft}, while compact, are not fully explicit. This is because these soft limit expressions involve field-space $s$-point correlation functions of the soft momenta, $F_{A_1,...,A_s}(\vect{p_1},...,\vect{p_s})$, (for the definition see Eq.~\eref{Ffieldspace}) corresponding to the white vertices in the diagrams. These $s$-point functions need to be evaluated at the time the last mode exits, $t_n$. However, explicit analytic expressions for field-space correlation functions are usually calculated when the evaluation time matches the earlier exit time of the $\vect{p_1},...,\vect{p_s}$. 

In this section we will calculate $F_{A_1,...,A_s}(\vect{p_1},...,\vect{p_s})$ in terms of  correlation functions evaluated at the earlier time at which the soft modes exit, for which there are analytic expressions available. To do so we will need to account for the evolution of field space perturbations themselves between successive flat hypersurfaces. 
This can be achieved by use of a separate universe expansion, analogous to the $\delta N$ expansion, which allows us to account for the evolution of field fluctuations between
horizon crossing times, given by the $\Gamma$ expansion
\begin{align}
\begin{split}
\delta \phi^{(l)}_{A,\vect{k}} =  \Gamma^{(le)}_{A,B} \delta \phi^{(e)}_{B,\vect{k }}  &+ \frac{1}{2}\Gamma^{(lee)}_{A,BC}\left [\delta \phi^{(e)}_{B}\star\delta \phi^{(e)}_{C}\right ]_{\vect{k }}+ \dots\,,
\\
\textrm{where }~~
\Gamma^{(le)}_{A,B}  \equiv &\frac{\partial \phi^{(l)}_{A}}{\partial \phi^{(e)}_{B}}\,, ~~
\Gamma^{(lee)}_{A,BC}  \equiv \frac{\partial^2 \phi^{(l)}_{A}}{\partial \phi^{(e)}_{B}\partial \phi^{(e)}_{C}}\,,
\end{split}
\label{gammaresummlater}
\end{align}
which expresses the perturbations on flat hypersurfaces at some later time $t_l$ in 
terms of the perturbation at some earlier 
time $t_e$. This was used at first-order in \cite{1507.08629} and to second-order in \cite{1511.03129}, and objects similar to the $\Gamma$ matrices have been used by a number of authors in the past\footnote{One can generalise the $\Gamma$ matrices to be $k$-dependent, as in \cite{Mulryne:2013uka}.}
\cite{Yokoyama:2007uu,Yokoyama:2007dw,Yokoyama:2008by,Avgoustidis:2011em,2011JCAP...11..005E,Seery:2012vj,Anderson:2012em,Elliston:2012ab}. We note that formally this expansion could be extended to include 
other degrees of freedom, just as we argued for the $\delta N$ and Soft Limit Expansion presented in 
earlier sections, though here we 
will only consider field perturbations. 

The real power of the $\Gamma$ matrix expansion
is that following a separate universe approach, 
they can be calculated with knowledge only of the 
background cosmology, in the same way that the $\delta N$ coefficients can also be calculated using just the background cosmology. This is what allows the correlations generated in specific models to be explicitly calculated. 
In Ref.~\cite{1507.08629}, for example, we gave explicit expressions for $\Gamma_{A,B}$ in 
canonical slow-roll models with 
sum-separable potentials, and these can easily be extended to second order.
Finally, we note that for the bispectrum and trispectrum, it will be sufficient to keep terms in the 
expansion up to first- and second-order, respectively, while we note that for $s$-point field-space 
correlation functions, the $(s-1)$th order contributions are required. 

In 
following sections we will make use of the result 
\begin{align}
N^{(e)}_{B } = N^{(l)}_{A }\Gamma^{(l,e)}_{A,B} \label{Ngamma}
\end{align}
which relates the earlier derivative of $N$ to the later one.

\subsection{Field-Space Correlation Functions}\label{sec:gammacor}

We can insert Eq.~\eref{gammaresummlater} into the field-space correlation functions $F_{A_1,...,A_s}(\vect{p_1},...,\vect{p_s})$ to express them in terms of correlation functions evaluated instead at the earlier times at which the soft modes exit the horizon, for which there are analytic expressions available.

We first consider the two-point function, evaluated at the late time $t_n$, for some soft momentum $\vect{p_1}$ which exits at an earlier time $t_1 < t_n$. Inserting the expansion Eq.~\eref{gammaresummlater} and taking the two-point function gives the tree level contribution (here, for clarity, we explicitly state the time label $(n)$, which was suppressed in previous sections)
\begin{align}
\Sigma^{(n)}_{AB}(p_1) = \Gamma^{(n,1)}_{A,C}\Gamma^{(n,1)}_{B,D}\Sigma^{(1)}_{CD}(p_1).
\label{sigmalate}
\end{align}
Note that the LHS of Eq.~\eref{sigmalate} was the object that appeared in the expressions of the previous section, such as Eq.~\eref{singlesoft3pt}, Eq.~\eref{internalsoft4pt}, Eq.~\eref{singex}, Eq.~\eref{kitetri}, Eq.~\eref{squished2} and Eq.~\eref{wonkytri}. 
The RHS of Eq.~\eref{sigmalate} involves the $\Gamma$ matrices -- set by the background cosmology -- and the field-space two-point function of the soft momentum at the time of horizon exit, $t_1$, which, specialising to canonical light fields,
 has the well known expression\footnote{
Strictly, this is the result that the two-point function takes 
after the decaying mode, present at horizon 
crossing, has decayed, written in terms of horizon crossing 
parameters.}  \cite{Stewart:1993bc,Nakamura:1996da} \begin{align}
\label{sigmaexit1}
\Sigma_{CD}^{(1)}(p_1) = \frac{{H^{(1)}}^2}{2 p_1^3}\delta_{CD}.
\end{align}

For the trispectrum in the double-soft-kite limit, Eq.~\eref{kitetri}, we also need the field space three-point function of soft momenta $\vect{p_1},\vect{p_2},\vect{p_3}$, evaluated at the later time $t_n$. Inserting three copies of Eq.~\eref{gammaresummlater} into the three-point function gives the tree level contribution \cite{Anderson:2012em,Seery:2012vj}
\begin{align}
\begin{split}
\alpha^{(n)}_{ABC}(p_1,p_2,p_3) = &\Gamma^{(n,1)}_{A,D}\Gamma^{(n,1)}_{B,E}\Gamma^{(n,1)}_{C,F}\alpha^{(1)}_{DEF}(p_1,p_2,p_3) 
\\
&+
\left [ \Gamma^{(n,11)}_{A,DE}\Gamma^{(n,1)}_{B,F}\Gamma^{(n,1)}_{C,G}\Sigma^{(1)}_{DF}(p_2)\Sigma^{(1)}_{EG}(p_3) + \left ( A,p_1 \to B,p_2 \to C,p_3 \right ) \right ]
\end{split}\label{alphalate}
\end{align}
where the three permutations in the second line are formed by cycling through $A \to B \to C$ whilst simultaneously cycling the momenta $p_1 \to p_2 \to p_3$. The LHS of Eq.~\eref{alphalate} appears in Eq.~\eref{kitetri} with $p_1 = k_1, p_2 = k_2$ and $p_3 = k_{12}$. The RHS of Eq.~\eref{alphalate} involves the first and second order $\Gamma$ matrices -- set by the background cosmology -- as well as the two- and three-point functions of the soft momenta at the time of horizon exit of the soft momenta $t_1$. The two-point function is given by Eq.~\eref{sigmaexit1} and the three-point function, again for canonical light fields, is given by \cite{astro-ph/0506056}
\begin{align}
\begin{split}
&\alpha_{DEF}^{(1)}(p_1,p_2,p_3) 
\\
&= \frac{4\pi^4}{p_1^3p_2^3p_3^3}\left (\frac{{H^{(1)}}}{2\pi}\right )^4\sum_{\text{6 perms}}\frac{\dot{\phi}_D^{(1)}\delta_{EF}}{4{H^{(1)}}}\left (-3\frac{p_2^2p_3^2}{p_t} - \frac{p_2^2p_3^2}{p_t^2}\left (p_1+2p_3\right ) +\frac{1}{2}p_1^2 - p_1p_2^2 \right )
\end{split} \label{SLalpha}
\end{align}
for $p_t \equiv p_1+p_2+p_3$, and where the sum is over the six permutations of $(DEF)$ while simultaneously rearranging the momenta $p_1,p_2,p_3$ such that the relative positioning of the $p$'s is respected.

We could proceed to consider $s$-point functions rewritten in terms of correlations at horizon crossing in a similar way, keeping terms in the $\Gamma$ expansion up to the $(s-1)$th order, but for now we have what is required for the bispectrum and trispectrum. We give the method for calculating the $s$-point functions using diagrams in App.~\ref{app:gammadiag}. 

\subsection{Explicit Examples}
\label{sec:explicit}
Continuing to consider models with canonical light fields,
we will now give our final expressions for the examples considered 
in \S\ref{sec:soft}, using the results of \S\ref{sec:gammacor}. The results given 
in this section can be used to compare a given multifield model against data. 
We will show results for dimensionless versions of the correlation 
functions.

\subsubsection*{Bispectrum}

The bispectrum can be parametrized in terms of the dimensionless quantity, 
\begin{align}
\fnl (k_1,k_2,k_3)\equiv \frac{5}{6}\frac{B_{\zeta}(k_1,k_2,k_3)}{\left [P_{\zeta}(k_1)P_{\zeta}(k_2)+ P_{\zeta}(k_2)P_{\zeta}(k_3)+P_{\zeta}(k_3)P_{\zeta}(k_1)\right ]}
\end{align}
known as the reduced bispectrum. 

For the squeezed limit of the bispectrum, the result Eq.~\eref{singlesoft3pt} becomes
 \begin{align}
 \lim _{k_1 \ll k_2,k_3} B_{\zeta}(k_1,k_2,k_3) 
\approx \ &-N^{(3)}_{A}N^{(3)}_{E}N^{(3)}_{E} \Gamma^{(3,1)}_{A,C}\Gamma^{(3,1)}_{B,C} \frac{d \phi^{(3)}_B}{d N} \frac{{H^{(1)}}^2}{2k_1^3} \frac{ {{H^{(3)}}}^2}{2k_3^3}\nonumber
\\
&+2N^{(3)}_{A}N^{(3)}_{EB}N^{(3)}_{E}  \Gamma^{(3,1)}_{A,C}\Gamma^{(3,1)}_{B,C}\frac{{H^{(1)}}^2}{2k_1^3}\frac{{{H^{(3)}}}^2}{2k_3^3} \label{bispectrumsqueezed2}
\end{align}
where we used Eq.~\eref{sigmalate} and Eq.~\eref{sigmaexit1} for $\Sigma_{AB}(k_1)$, and we differentiated the $\delta N$ expression for $P_{\zeta } (k_3)$.

If the bispectrum is large enough to be observed 
by present or near future probes, the second term in  Eq.~\eref{bispectrumsqueezed2}
must be dominant, and we can then form the dimensionless reduced bispectrum 
\begin{align}
\lim_{k_1 \ll k_2,k_3} \fnl \equiv \frac{5}{12}\frac{\lim\limits_{k_1 \ll k_2,k_3}B_{\zeta}(k_1,k_2,k_3)}{P_{\zeta}(k_1)P_{\zeta}(k_3)} & \approx  \frac{5}{6}
\frac{N^{(3)}_{A}N^{(3)}_{EB}N^{(3)}_{E}  \Gamma^{(3,1)}_{A,C}\Gamma^{(3,1)}_{B,C}  }
{N^{(3)}_{D}N^{(3)}_{D}N^{(3)}_{F}N^{(3)}_{G}
   \Gamma^{(3,1)}_{F,H}\Gamma^{(3,1)}_{G,H} }       \,,
   \end{align}
which is dependent on the two scales $k_1$ and $k_3$ through the two horizon crossing times.

We could use Eq.~\eref{Ngamma} to write this more succinctly as
\begin{align}
\lim_{k_1 \ll k_2,k_3} \fnl \approx  \frac{5}{6}
\frac{N^{(1)}_{A}N^{(3)}_{BC}N^{(3)}_{B}  \Gamma^{(3,1)}_{C,A}  }
{N^{(3)}_{D}N^{(3)}_{D}N^{(1)}_{E}N^{(1)}_{E}
     }       \,,\label{fnlsqueeze}
   \end{align}   
which can be contrasted with the usual $\delta N$ formula for the reduced bispectrum,
valid for close to equilateral configurations (under the same assumptions)
\begin{align}
\lim_{k_1 \approx k_2\approx k_3}\fnl \equiv \frac{5}{6}\frac{\lim\limits_{k_1 \approx k_2 \approx k_3}B_{\zeta}(k_1,k_2,k_3)}{\left [P_{\zeta}(k_1)P_{\zeta}(k_2) + 2 \ \rm perms\right ]} = \frac{5}{6}
\frac{N^{(3)}_{A}N^{(3)}_{AB}N^{(3)}_{B}    }
{N^{(3)}_{C}N^{(3)}_{C}N^{(3)}_{D}N^{(3)}_{D}
    }\,,\label{fnlequil}
\end{align}
which retains dependence on only 
a single horizon crossing time. In \cite{1507.08629} we found that the difference between 
Eq.~\eref{fnlsqueeze} and Eq.~\eref{fnlequil} can be very important for models with significant scale dependence. For this we looked at a simple two-field example of a mixed inflaton-curvaton model with curvaton self-interactions. For that model it was possible to derive analytic expressions for the derivatives of $N$ and the first-order $\Gamma$ matrices. We expect it to be possible to calculate higher-order $\Gamma$ matrices in this model as well, but we leave this for future work.

\subsubsection*{Trispectrum}

Turning to the trispectrum, let us first review results for the case in 
which all modes 
cross the horizon at close to the same time, if the trispectrum is observable by present 
or near future observations then in this case it is given by \cite{Seery:2006vu,Byrnes:2006vq}
\begin{align}
\begin{split}
T_{\zeta}(\vect{k_1}, \vect{k_2}, \vect{k_3}, \vect{k_4}) = &\tnl\left [P_{\zeta}(k_{13})P_{\zeta}(k_3)P_{\zeta}(k_4)+ \rm \ 11 perms\right ] 
\label{locTri}
\\
&+ \gnl\left [P_{\zeta}(k_{2})P_{\zeta}(k_3)P_{\zeta}(k_4)+ \rm 4 \  perms\right ]
\end{split}
\end{align}
where
\begin{align}
 {\tau}_{\rm NL}& \equiv \frac{N^{(3)}_{AB}N^{(3)}_BN^{(3)}_{AC}N^{(3)}_C }{(N^{(3)}_D N^{(3)}_D)^3} \label{tnl}
 \\
  {g}_{\rm NL}&\equiv \frac{N^{(3)}_{ABC}N^{(3)}_AN^{(3)}_{B}N^{(3)}_C }{(N^{(3)}_D N^{(3)}_D)^3} \label{gnl}
\end{align}
are dimensionless parameters which represent the amplitude 
of two distinct shapes.

This decomposition into two shapes is 
only useful for close to equilateral configurations.
For soft limits one can, however, form dimensionless versions of the trispectrum. In the case of the single-soft limit of the trispectrum, we will form the single-soft dimensionless trispectrum by dividing the trispectrum by one power spectrum evaluated on the soft momentum, and two copies of the power spectrum evaluated on the hard momentum. For the double-soft limit, we will form the double-soft dimensionless trispectrum by dividing the trispectrum by two copies of the power spectrum evaluated on the soft momentum, and one power spectrum evaluated on the hard momentum.

\subsubsection*{Single-Soft Squeezed Trispectrum}

For the single-soft squeezed limit of the trispectrum, we obtained the result Eq.~\eref{singex}, which contains the derivative of the near-equilateral bispectrum, $B_{\zeta}(k_2,k_3,k_4)$. The 
standard $\delta N$ expression in near equilateral configurations 
can be used to give an expression for this piece, which we then need to differentiate. 
Focusing again on the case in which the bispectrum and trispectrum are observably large, and 
using Eq.~\eref{sigmalate} to write $\Sigma_{AB}(k_1)$ in terms of known quantities and $\Gamma$ matrices
we find
\begin{align}
f^{\rm ext}_4(\vect{k_1}, \vect{k_2}, \vect{k_3}, \vect{k_4}) &\equiv  
\frac{\lim\limits_{k_1 \ll k_2 \approx k_3 \approx k_4 } T_{\zeta}(\vect{k_1}, \vect{k_2}, \vect{k_3}, \vect{k_4})}{3P_{\zeta}(k_1)\left [P_{\zeta}(k_4)\right ]^2}
\\
&\approx 	\frac{N^{(4)}_A\Gamma^{(4,1)}_{A,E}\Gamma^{(4,1)}_{B,E}  N^{(4)}_{CDB}N^{(4)}_CN^{(4)}_D}
{N^{(4)}_{G}N^{(4)}_{H}\Gamma^{(4,1)}_{G,F}\Gamma^{(4,1)}_{H,F}(N^{(4)}_{I}N^{(4)}_{I})^2} + 	\frac{N^{(4)}_A\Gamma^{(4,1)}_{A,E}\Gamma^{(4,1)}_{B,E}   N^{(4)}_{CD}N^{(4)}_{CB}N^{(4)}_D}{N^{(4)}_{G}N^{(4)}_{H}\Gamma^{(4,1)}_{G,F}\Gamma^{(4,1)}_{H,F}(N^{(4)}_{I}N^{(4)}_{I})^2}.
\end{align}
We can again use Eq.~\eref{Ngamma} to write this more succinctly as
\begin{align}
\label{singleSoft}
f^{\rm ext}_4(\vect{k_1}, \vect{k_2}, \vect{k_3}, \vect{k_4})
\approx 	
\frac{N^{(1)}_A \Gamma^{(4,1)}_{B,A}  N^{(4)}_{CDB}N^{(4)}_CN^{(4)}_D}
{N^{(1)}_{E}N^{(1)}_{E}(N^{(4)}_{F}N^{(4)}_{F})^2}
+ 	
\frac{N^{(1)}_A\Gamma^{(4,1)}_{B,A}   N^{(4)}_{CD}N^{(4)}_{CB}N^{(4)}_D}{N^{(1)}_{E}N^{(1)}_{E}(N^{(4)}_{F}N^{(4)}_{F})^2}.
\end{align}

If we had taken the near-equilateral configuration, Eq.~\eref{locTri}, and formed the analogous reduced trispectrum and pushed 
this expression towards the 
single-soft squeezed limit,  we would have found a contribution both from the $\tau_{\rm NL}$ and the $g_{\rm NL}$ 
shape in this limit, and we would have arrived at a similar expression to Eq.~\eref{singleSoft} but with all the $N$ derivatives appearing 
with the same superscript and $\Gamma^{(4,1)}_{A,B}$
replaced with $\delta_{AB}$. Our expression can be significantly different from this naive one.
\

\subsubsection*{Single-Soft Internal Trispectrum}
We now proceed to produce similar expressions for the other soft limits.
For the single-soft collapsed limit of the trispectrum, we obtained the result Eq.~\eref{internalsoft4pt}. 
The derivative of the power spectrum gives
\begin{align}
P_{\zeta}(k_1) _{,A} = 2 N_{BA} N_C \Sigma_{BC} (k_1) +  N _B N _C \Sigma_{BC,A} (k_1)\,.
\label{powerspecderiv}
\end{align}
Utilising this expression, keeping only the first term, we find
\begin{align}
f^{\rm int}_4(\vect{k_1}, \vect{k_2}, \vect{k_3}, \vect{k_4}) \equiv \frac{\lim\limits_{k_{12} \ll k_1 \approx k_2 \approx k_3 \approx k_4 } T_{\zeta}(\vect{k_1}, \vect{k_2}, \vect{k_3}, \vect{k_4})}{4P_{\zeta}(k_{12})\left [P_{\zeta}(k_4)\right ]^2} \approx \frac{N^{(4)}_{AB}N^{(4)}_{A}\Gamma^{(4,{12})}_{B,C}\Gamma^{(4,{12})}_{D,C}N^{(4)}_{ED}N^{(4)}_{E}}{N^{(12)}_{F}N^{(12)}_{F}(N^{(4)}_{G}N^{(4)}_{G})^2}. \label{singlesoftintexpl}
\end{align}

Considering Eq.~\eref{locTri} for close to equilateral configurations and pushing this expression towards 
the collapsed limit, we could have found a contribution from the $\tau_{\rm NL}$ shape 
alone, and a similar expression to Eq.~\eref{singlesoftintexpl}, but with all the $N$ derivatives appearing 
with the same superscript and $\Gamma^{(4,{12})}_{A,B}$
replaced with $\delta_{AB}$. Once again our new expression can be significantly altered from this naive expression.

\subsubsection*{Double-Soft Trispectrum} 

\

\textbf{1. Kite}

We had Eq.~\eref{kitetri}
from the soft diagrams, so here we can once again use Eq.~\eref{sigmalate} and Eq.~\eref{sigmaexit1} to write $\Sigma_{AB}(k_1)$ and $\Sigma_{AB}(k_2)$ in terms of known quantities and $\Gamma$ matrices, and now we also need to use Eq.~\eref{alphalate} with Eq.~\eref{SLalpha} and Eq.~\eref{sigmaexit1} to write $\alpha^{(4)}_{ABC}(k_1,k_2,k_{12})$, in 
terms of horizon crossing expressions 
and $\Gamma$ matrices -- in this case the second order $\Gamma$ coefficient is necessary. 
We note that there is nothing to stop $\alpha_{ABC}(k_1,k_2,k_{12})$ from becoming large even in canonical models, 
because it has evolved from time $t_1$ to time $t_4$. 

Keeping all the terms that can be significant and taking $k_s \approx k_1 \approx k_2 \approx k_{12}$ and $k_h \approx k_3 \approx k_4$ with $k_s \ll k_h$ to define the kite limit, then the dimensionless quantity  appropriate here is
\begin{align}
&f^{\rm kite}_4(\vect{k_1}, \vect{k_2}, \vect{k_3}, \vect{k_4}) \equiv \frac{\lim\limits_{ \rm kite } T_{\zeta}(\vect{k_1}, \vect{k_2}, \vect{k_3}, \vect{k_4})}{2\left [P_{\zeta}(k_s)\right ]^2P_{\zeta}(k_h)} 
\\
\begin{split}
&\approx 
\frac{N^{(h)}_{A}  N^{(h)}_{B}  N^{(h)}_{FC}  N^{(h)}_{F}  
\Gamma^{(h,s)}_{A,DE} \Gamma^{(h,s)}_{B,D} \Gamma^{(h,s)}_{C,E} }
{\left (N^{(s)}_{G}N^{(s)}_{G}\right )^2N^{(h)}_{H}N^{(h)}_{H}} 
\\
&+ 
\frac{N^{(h)}_{A}  N^{(h)}_{B}  \left (N^{(h)}_{ECD}  N^{(h)}_{E} + N^{(h)}_{EC}  N^{(h)}_{DE} \right )
\Gamma^{(h,s)}_{A,F} \Gamma^{(h,s)}_{C,F} \Gamma^{(h,s)}_{B,I} \Gamma^{(h,s)}_{D,I}}
{\left (N^{(s)}_{G}N^{(s)}_{G}\right )^2N^{(h)}_{H}N^{(h)}_{H}} 
\\
&+ 
\frac{2N^{(h)}_{A}  N^{(h)}_{BC}  N^{(h)}_{E}  N^{(h)}_{ED}  
\Gamma^{(h,s)}_{A,F} \Gamma^{(h,s)}_{B,F} \Gamma^{(h,s)}_{C,I} \Gamma^{(h,s)}_{D,I}}
{\left (N^{(s)}_{G}N^{(s)}_{G}\right )^2N^{(h)}_{H}N^{(h)}_{H}} 
\end{split}
\label{doublesoftkiteexp}
\end{align}

We note once again that considering Eq.~\eref{locTri} for close to equilateral configurations and pushing 
this towards the kite limit, both the $\tau_{\rm NL}$ and $g_{\rm NL}$ shapes would have contributed in this limit and 
we would have found similar terms to the last two lines of Eq.~\eref{doublesoftkiteexp}.
The term in the first line of Eq.~\eref{doublesoftkiteexp}, however, is qualitatively different, and we believe is a new form of possibly significant contribution to the trispectrum, which 
could be large, even in models where the equilateral $\gnl$ and $\tnl$ are small. We hope to investigate this 
further in future work.

\

\textbf{2. Squished}

The squished limit gave the same expression Eq.~\eref{squished2} as the single-soft collapsed  limit Eq.~\eref{internalsoft4pt}. We refer to the corresponding result Eq.~\eref{singlesoftintexpl} for the explicit form of the reduced trispectrum in this case. 

\

\textbf{3. Wonky}

For this limit we had Eq.~\eref{wonkytri}.
The dimensionless trispectrum relevant here is
\begin{align}
f^{\rm wonky}_4(\vect{k_1}, \vect{k_2}, \vect{k_3}, \vect{k_4}) &\equiv \frac{\lim\limits_{ k_1 \ll k_2 \ll k_3 \approx k_4 } T_{\zeta}(\vect{k_1}, \vect{k_2}, \vect{k_3}, \vect{k_4})}{2P_{\zeta}(k_{1})P_{\zeta}(k_{2}) P_{\zeta}(k_4)} 
\\
&\approx \frac{N^{(4)}_{A}\Gamma^{(4,1)}_{A,E}\Gamma^{(4,1)}_{B,E}\left (N^{(4)}_{C}\Gamma^{(4,2)}_{C,F}\Gamma^{(4,2)}_{D,F}N^{(4)}_{GD}N^{(4)}_{G} \right )_{,B}}{N^{(4)}_{F}N^{(4)}_{F}N^{(1)}_{G}N^{(1)}_{G} N^{(2)}_{H}N^{(2)}_{H}} \label{wonky1}
\end{align}
where we have again neglected both first and second derivatives of $\Sigma^{ (s)}_{AB}(k_s)$. Note the 
comments concerning the kite limit, and a new contribution to the trispectrum, also apply here.

\section{Conclusion}
\label{conc}

In this paper we developed a formalism for calculating soft limits of $n$-point inflationary correlation functions for multiple light fields. This formalism allows for squeezed (external) or collapsed (internal) soft modes and for multiple soft modes either of the same size or with a hierarchy amongst the soft modes. We used a diagrammatic approach to organise the separate universe Soft Limit Expansion Eq.~\eref{softexpansion1} to allow for explicit computation of any $n$-point $\zeta$ correlators in soft limit polygon shapes.

We applied our results to derive new, explicit expressions for the single- and double-soft limits of the trispectrum for a variety of quadrilateral shapes. A highlight of this section was the identification of possibly large contributions 
to the trispectrum in canonical models which are missed by the usual analysis. 
We also gave a new, direct proof of the soft limit version of the Suyama-Yamaguchi inequality and generalized this to give an infinite tower of new inequalities between soft limits of $n$-point correlators which are constrained by products of $2r$- and $2(n-r)$-point equilateral correlators, with $ 1 \leq r  \leq n-2$. The case of $n=3, r=2$ is the well-known Suyama-Yamaguchi inequality --  with other choices of $n$ and $r$ representing new inequalities. All of these are saturated in single-source models, and their violation may signify a breakdown of the inflationary paradigm.

We emphasize that these results are important for future observations which can probe a larger range of scales. Accurate theoretical predictions may be necessary, even for the trispectrum, and we may hope to rule out large classes of inflationary models by observations of these soft limits. For example, DES, Euclid and $\mu$-distortion experiments will all provide new constraints in the near future.

\section*{Acknowledgements}
We would like to thank David Seery for the idea of expanding the hard $\zeta$ mode directly 
in terms of the soft field fluctuations, which greatly simplifies our methodology. We would like to thank Chris Byrnes, Shailee Imrith and Enrico Pajer for helpful feedback and insightful comments on a draft of this paper.
 ZK is supported by an STFC studentship. 
DJM is supported by a Royal Society University Research Fellowship.

\appendix

\section{The background wave method}
\label{appBwave}
In this appendix we review the background wave method as used previously for models of single-field inflation (see e.g. \cite{Maldacena:2002vr,Creminelli:2004yq,Cheung:2007sv,Senatore:2012wy}) and then show how this generalises to models of multi-field inflation. In summary, for single-field inflation, the background wave method allows the effect of a soft mode to be traded for a change of spatial coordinates. In multi-field inflation however, one can't trade for a change in spatial coordinates, but instead one can trade for a change in field-space background values.

We begin with how soft $\zeta$ modes can be set to zero, if one performs a suitable change of spatial coordinates.
Since $\zeta$ appears in the perturbed spatial metric as $g_{ij}(t,\vect{x})dx^idx^j= a^2(t)e^{2\zeta(t,\vect{x})}\delta_{ij}dx^idx^j$, a soft mode $\zeta^{\rm s}(t,\vect{x})$ can be recast as a change in coordinates, i.e. one can set $\zeta^{\rm s}(t,\vect{x}) \mapsto 0$ as long as one performs the coordinate transformation \cite{Cheung:2007sv}
\begin{align}
\vect{x} \mapsto \vect{x}' \equiv e^{\zeta^{\rm s}(t,\vect{x})}\vect{x} \label{zetacoord}
\end{align}
everywhere. This statement is true of both single- and muti-field inflation. However, its 
utility for application to soft limits depends on whether one works with a single- or multi-field inflation model.

For single-field inflation, the method runs as follows: in single-field inflation, one assumes that the hard $\zeta$ mode\footnote{or correlation functions of hard $\zeta$ modes} in the presence of soft modes, denoted $\zeta^{\rm h}(\vect{x})\Big|_{ \rm s}$,   feels the effect of the soft modes \textit{only} through dependence on the soft $\zeta$ modes 
\begin{align}
\zeta^{\rm h}(\vect{x})\Big|_{ \rm s} = \zeta^{\rm h}(\vect{x})\Big|_{ \zeta^{\rm s}}. \label{singlefieldsoftzeta}
\end{align}
Note that equation Eq.~\eref{singlefieldsoftzeta} holds only for single-field inflation since in this case there are no isocurvature modes, $\chi_\alpha$, and hence no dependence on soft isocurvature modes, $\chi_\alpha^{\rm s}$. The relation Eq.~\eref{zetacoord} means that $\zeta^{\rm h}(\vect{x})\Big|_{ \zeta^{\rm s}}  = \zeta^{\rm h}(\vect{x}')\Big|_{0}$, where the subscript zero indicates the value the hard modes take in the absence of any soft modes. This can then be inserted into Eq.~\eref{singlefieldsoftzeta} to give, for single-field inflation,
\begin{align}
\zeta^{\rm h}(\vect{x})\Big|_{ \rm s} = \zeta^{\rm h}(\vect{x}')\Big|_{0}.\label{singlefieldcoor}
\end{align}
The interpretation of this is that for single-field models, the effect of soft modes on the hard $\zeta$ can be accounted for by just rescaling the spatial coordinates and evaluating the hard mode in the absence of any soft modes. One can then Taylor expand the RHS of Eq.~\eref{singlefieldcoor}, for small $\zeta^{\rm s}$, to get (see e.g. \cite{Creminelli:2004yq,Cheung:2007sv})
\begin{align}
\zeta^{\rm h}(\vect{x}')\Big|_{0} = \zeta^{\rm h}(\vect{x})\Big|_{0} + \zeta^{\rm s} \ \vect{x}\cdot \nabla\zeta^{\rm h}(\vect{x})\Big|_{0} + \cdots\label{sfzetaex}
\end{align}
which holds for single-field models. This can then be inserted into soft correlation functions to derive single-field consistency relations.

For multi-field models, Eq.~\eref{zetacoord} still holds, but Eq.~\eref{singlefieldsoftzeta} does not. This means one doesn't have Eq.~\eref{singlefieldcoor} or Eq.~\eref{sfzetaex}. Instead of Eq.~\eref{singlefieldsoftzeta}, for multi-field models we assume
\begin{align}
\zeta^{\rm h}(\vect{x})\Big|_{ \rm s} = \zeta^{\rm h}(\vect{x})\Big|_{ \zeta^{\rm s}, \chi_\alpha^{\rm s}}
\end{align}
where the $\chi_\alpha^{\rm s}$ are soft isocurvature modes. Equivalently, since $\zeta$ and $\chi_\alpha$ can be recast in terms of fluctuations of the multiple scalar fields, we have
\begin{align}
\zeta^{\rm h}(\vect{x})\Big|_{ \rm s} = \zeta^{\rm h}(\vect{x})\Big|_{ \delta\phi_A^{\rm s}}\label{mfsoft}
\end{align}
The interpretation of Eq.~\eref{mfsoft} is that for multi-field inflation, the effect of soft modes on the hard $\zeta$ is not just a local rescaling of coordinates (which was the single-field case), but is instead a more general transformation in the background values of the multiple scalar field values. 

When inserted into soft limits of correlation functions, Eq.~\eref{mfsoft} implies that we are assuming that the main contribution to correlations between hard and soft modes comes from how the soft modes, which exit the horizon at much earlier times, alter the background cosmology in which hard modes exit. This assumption can be used for any set of scales, but becomes accurate only when the hierarchy is large. We then can Taylor expand the RHS of Eq.~\eref{mfsoft} in powers of $\delta\phi_A^{\rm s}$ around the value it would have taken in the absence of any soft scale modes $\zeta^{\rm h}(\vect{x})\Big|_{0} \equiv \zeta^{\rm h}(\vect{x})$ to get Eq.~\eref{softexpansion1}.

\section{Reduction to Single Field Result}
\label{app:single}
We would like to recover the single-field result (5) of \cite{Joyce:2014aqa} (which agrees with the results of \cite{Mirbabayi:2014zpa}) for the double-soft kite limit, with their $N$ set to $N=2$
\begin{align}
\lim _{\rm k_1 \approx k_2 \ll k_3 \approx k_4} T_{\zeta}(\vect{k_1}, \vect{k_2}, \vect{k_3}, \vect{k_4}) &= B_{\zeta}(k_1,k_2,k_{12})\delta_{\mathcal{D}}P_{\zeta}(k_3) + P_{\zeta}(k_1)P_{\zeta}(k_2)\delta_{\mathcal{D}}^2P_{\zeta}(k_3)
\\
\textrm{where }\delta_{\mathcal{D}} &= -3 - \frac{d}{d \log k_3}.
\end{align}
In the single field case, our expression Eq.~\eref{kitetri} reduces to
\begin{align}
\begin{split}
&\lim _{\rm k_1 \approx k_2 \ll k_3 \approx k_4} T_{\zeta}(\vect{k_1}, \vect{k_2}, \vect{k_3}, \vect{k_4}) 
\\ = &N^2_{\phi}P_{\zeta}(k_3)_{,\phi}\alpha_{\phi\phi\phi}(k_1,k_2,k_{12}) + N^2_{\phi}P_{\zeta}(k_3)_{,\phi\phi}\Sigma_{\phi\phi}(k_1)\Sigma_{\phi\phi}(k_2)
\\
+& N_{\phi}N_{\phi\phi}P_{\zeta}(k_3)_{,\phi}\Sigma_{\phi\phi}(k_1)\Sigma_{\phi\phi}(k_{12})+ N_{\phi}N_{\phi\phi}P_{\zeta}(k_3)_{,\phi}\Sigma_{\phi\phi}(k_2)\Sigma_{\phi\phi}(k_{12}).
\end{split} \label{singlefieldred}
\end{align}
We will use the standard single-field slow-roll expressions $N_{\phi}=-H/\dot{\phi} = 1/\sqrt{2\epsilon_V}$ and $N_{\phi\phi} = 1 -\eta_V/(2\epsilon_V)$ where $\epsilon_V \equiv \frac{1}{2}\left (V'/V\right )^2$ and $\eta_V \equiv  V''/V   $ are the potential slow roll parameters for slow-roll potential $V$. We will also use $\Sigma(k_3)_{\phi\phi}= {H^2}/{2k_3^3}$. Note that the scalar spectral index is 
\begin{align}
n_s-1 = 2\eta_V - 6\epsilon_V = -\frac{(1+2N_{\phi\phi})}{N_{\phi}^2} \label{singlefieldns}
\end{align}
and in terms of the dilatation operator we have 
\begin{align}
n_s-1 =-\frac{1}{P_{\zeta}(k_3)}\delta_{\mathcal{D}}P_{\zeta}(k_3). \label{singlefieldnsdil}
\end{align}
The scalar tilt is given by
\begin{align}
\alpha_s \equiv \frac{d\log (n_s-1)}{d \log k} = -\frac{(n_s-1)_{,\phi}}{N_\phi(n_s-1)}\label{alphas}
\end{align}
where we've used $k=aH$ at horizon exit and $N_{\phi}=-H/\dot{\phi}$.

We would like to have expressions for $P_{\zeta}(k_3)_{,\phi}$ and $P_{\zeta}(k_3)_{,\phi\phi}$. Firstly
\begin{align}
P_{\zeta}(k_3)_{,\phi} = \left (N_{\phi}^2\frac{H^2}{2k_3^3}\right )_{,\phi} = \frac{P_{\zeta}(k_3)}{N_{\phi}}(1+2N_{\phi\phi}) =-N_{\phi}(n_s-1)P_{\zeta}(k_3) = N_{\phi}\delta_{\mathcal{D}}P_{\zeta}(k_3) \label{pphi}
\end{align}
where in the second equality we have used the slow-roll result $\left (H^2\right )_{,\phi} = H^2/N_{\phi}$, in the third equality we have used Eq.~\eref{singlefieldns} and in the final equality we used Eq.~\eref{singlefieldnsdil}. Next, the second derivative is 

\begin{align}
P_{\zeta}(k_3)_{,\phi\phi} &= \left (-N_{\phi}(n_s-1)P_{\zeta}(k_3)\right )_{,\phi} 
\\
&= -N_{\phi\phi}(n_s-1)P_{\zeta}(k_3) + N_{\phi}\left [\alpha_s + N_{\phi}(n_s-1) \right ](n_s-1)P_{\zeta}(k_3) 
\\
&= N_{\phi\phi}\delta_{\mathcal{D}}P_{\zeta}(k_3) + N_{\phi}\left [-3 - \frac{d}{d \log k_3} \right ]\left (-(n_s-1)P_{\zeta}(k_3) \right )
\\
&= N_{\phi\phi}\delta_{\mathcal{D}}P_{\zeta}(k_3) + N_{\phi}\delta_{\mathcal{D}}^2P_{\zeta}(k_3)\label{pphiphi}
\end{align}
where we've made repeated use of Eq.~\eref{pphi}, together with Eq.~\eref{alphas} in the second line. We can now substitute Eq.~\eref{pphi} and Eq.~\eref{pphiphi} into Eq.~\eref{singlefieldred} and factor out $\delta_{\mathcal{D}}P_{\zeta}$ and $\delta_{\mathcal{D}}^2P_{\zeta}$ terms  to give
\begin{align}
\begin{split}
&\lim _{\rm k_1 \approx k_2 \ll k_3 \approx k_4} T_{\zeta}(\vect{k_1}, \vect{k_2}, \vect{k_3}, \vect{k_4}) 
\\ = &\left \{N_{\phi}^3\alpha_{\phi\phi\phi}(k_1,k_2,k_{12}) + N_{\phi}^2N_{\phi\phi}\left [\Sigma_{\phi\phi}(k_1)\Sigma_{\phi\phi}(k_2) +(k_1 \to k_2 \to k_{12})  \right ]\right \}\delta_{\mathcal{D}}P_{\zeta}(k_3)
\\
&+ N_{\phi}^4\Sigma_{\phi\phi}(k_1)\Sigma_{\phi\phi}(k_2)\delta_{\mathcal{D}}^2P_{\zeta}(k_3)
\end{split} 
\\
= & B_{\zeta}(k_1,k_2,k_{12})\delta_{\mathcal{D}}P_{\zeta}(k_3) + P_{\zeta}(k_1)P_{\zeta}(k_2)\delta_{\mathcal{D}}^2P_{\zeta}(k_3)
\end{align}
which is the desired single field result.

\section{$\Gamma$ Diagrams}
\label{app:gammadiag}
In \S\ref{sec:gamma} we made the $\Gamma$ expansion Eq.~\eref{gammaresummlater} expressing perturbation on flat hypersurfaces at some later time in terms of the perturbations on an earlier flat hypersurface. 
Then in \S\ref{sec:gammacor} we inserted this expansion into the late time field-space two- and three-point functions. 
In this appendix, we consider the more general case of the field-space $s$-point function $F^{(n)}_{A_1\cdot \cdot \cdot A_s}(\vect{p_1},...,\vect{p_s})$, defined in Eq.~\eref{Ffieldspace}, writing it in terms of field-space correlation functions whose evaluation time matches the exit time of the soft modes. We must expand up to $(s-1)$-th order to be consistent. Wick contractions then occur between the terms. Again,  we can organise the result in terms of diagrams, which we call the $\Gamma$ diagrams, which are analogous to both the $\delta N$ Diagrams and the Soft Limit Diagrams. As before, we focus on tree-level, and keep only leading order terms in the gradient expansion.

\

\textbf{$\Gamma$ Diagram Rules:}
\begin{enumerate}
\I Draw $s$-external dashed lines, labelled with incoming momenta $\vect{p_a}$ and field index $A_a$ for $a=1,...,s$. Draw a cross vertex at the end of each dashed line.
\I Connected the cross vertices by drawing a connected tree diagram with wavy lines. Each wavy line must connect on one end to a cross vertex
and on the other end to a square vertex. At a cross vertex, (possibly multiple) wavy lines can connect to a dashed line. At a square vertex wavy lines connect to other wavy lines. 
\I Label each wavy line with a distinct field index $B_1,B_2,...$.
\I Ensure momentum conservation at every vertex, which determines the momentum of each wavy line.
\I The two vertex types are assigned the following factors
\begin{enumerate}
\I Assign a factor $\Gamma^{(n,1...1)}_{A_a, B1 \cdot \cdot \cdot B_m}$ to each cross vertex with one external dashed line with index $A_a$ and $m$ wavy lines with field indices $B_1,\cdot \cdot \cdot ,B_m$, where $1 \leq m \leq s-1$. 
\I Assign a factor $F^{(1)}_{B_1\cdot \cdot \cdot B_r}(\vect{q_1},...,\vect{q_r})$ to each square vertex with no dashed external lines and $r$ wavy lines with incoming momenta $\vect{q_1},...,\vect{q_s}$ and field indices $B_1\cdot \cdot \cdot B_r$, where $2 \leq r \leq s$.
\end{enumerate} 
\I Each diagram is associated with the mathematical expression obtained by multiplying together all vertex factors.
Repeat the above process from stage 2 onwards to generate all distinct connected tree diagrams.
 $F^{(n)}_{A_1\cdot \cdot \cdot A_s}(\vect{p_1},...,\vect{p_s})$ is then obtained by summing over all these diagrams.
\end{enumerate}
 
\begin{figure}[htb!]
\center
\includegraphics[scale=0.48]{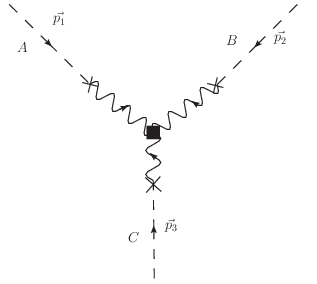} 
\includegraphics[scale=0.48]{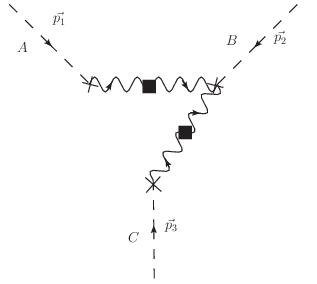} 
\includegraphics[scale=0.48]{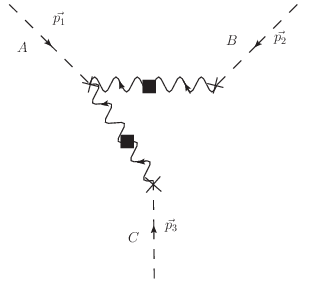} 
\includegraphics[scale=0.48]{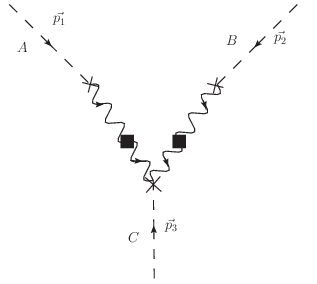} 
\caption{Distinct tree-level connected $\Gamma$ Diagrams for the field space three-point function}
\label{fig:gam_3}
\end{figure}

As an example, in Fig.~\ref{fig:gam_3}, we show the $\Gamma$ diagrams for the late-time field-space three-point function, corresponding to the expression Eq.~\eref{alphalate} \cite{1511.03129}, which we repeat here for convenience
\begin{align}
\begin{split}
\alpha^{(n)}_{ABC}(p_1,p_2,p_3) = &\Gamma^{(n,1)}_{A,D}\Gamma^{(n,1)}_{B,E}\Gamma^{(n,1)}_{C,F}\alpha^{(1)}_{DEF}(p_1,p_2,p_3) 
\\
&+
\left [ \Gamma^{(n,11)}_{A,DE}\Gamma^{(n,1)}_{B,F}\Gamma^{(n,1)}_{C,G}\Sigma^{(1)}_{DF}(p_2)\Sigma^{(1)}_{EG}(p_3) + \left ( A,p_1 \to B,p_2 \to C,p_3 \right ) \right ].
\end{split} 
\end{align}

\bibliographystyle{JHEP}

\bibliography{tri.bib}

\end{document}